\documentclass[12pt,a4paper]{article}
\pdfoutput=1
\usepackage{jheppub}
\usepackage{amsmath}
\usepackage{amsfonts}
\usepackage{mathtools}
\usepackage{amssymb}
\usepackage[utf8]{inputenc}
\allowdisplaybreaks[2]
\def\be{\begin{equation}}
\def\ee{\end{equation}}
\def\bea{\begin{eqnarray}}
\def\eea{\end{eqnarray}}
\let\fr=\frac
\def\W={\cal W}
\def\L ={\cal L}

\let\pa=\partial

\def\dalemb#1#2{{\vbox{\hrule height .#2pt
        \hbox{\vrule width.#2pt height#1pt \kern#1pt
                \vrule width.#2pt}
        \hrule height.#2pt}}}

\title{Entanglement of heavy quark impurities and generalized gravitational entropy }
\author{S. Prem Kumar} 
\author{and Dorian Silvani} 
\affiliation{Department of Physics,\\Swansea University, \\ 
Singleton Park,\\ Swansea, SA2 8PP, U.K.}
\emailAdd{s.p.kumar@swansea.ac.uk}
\emailAdd{d.silvani.492808@swansea.ac.uk}
\abstract{ 
We calculate the contribution from non-conformal heavy quark sources to the entanglement entropy (EE) of a spherical region in ${\cal N}=4$ SUSY Yang-Mills theory. We apply the generalized gravitational entropy method to non-conformal probe D-brane embeddings in AdS$_5\times$S$^5$, dual to pointlike impurities exhibiting flows between quarks in large-rank tensor representations and the fundamental representation. For the D5-brane embedding which describes the screening of fundamental quarks in the UV to the   antisymmetric tensor representation in the IR, the EE excess decreases non-monotonically towards its IR
asymptotic value,  tracking the qualitative  behaviour of the 
one-point function of static fields sourced by the impurity. We also examine two classes of D3-brane embeddings, one which connects a symmetric representation source in the UV to fundamental quarks in the IR, and a second category which yields the symmetric representation source on the Coulomb branch. The EE excess for the former increases from the UV to the IR, whilst decreasing and becoming negative for the latter.  In all cases, the probe free energy on hyperbolic space with $\beta=2\pi$ increases monotonically towards the IR, supporting its interpretation as a relative entropy. We identify universal corrections, depending logarithmically on the VEV, for the symmetric representation on the Coulomb branch.}
\begin{document}
\maketitle
\flushbottom
\section{Introduction}
The holographic correspondence \cite{maldacena, witten, gkp} between gauge theories and gravity has revealed an intriguing link between quantum entanglement and geometry \cite{Ryu:2006bv, Ryu:2006ef, Casini:2011kv, Maldacena:2013xja}. The prescription of  \cite{Ryu:2006bv, Ryu:2006ef, Casini:2011kv} relating  the entanglement entropy of some subsystem within a quantum system  to the area of an extremal surface in a classical dual gravity framework, was put on firm footing in \cite{Lewkowycz:2013nqa}, where the replica trick was implemented in the gravity setting dual to the subsystem of interest, by  using the method of  \cite{Callan:1994py}. This involves identifying a circle in the asymptotic geometry, which could be a compact Euclidean time direction, varying its periodicity in a well-defined manner and calculating the resulting variation in the action so as to obtain a gravitational or geometric entropy.  

A natural extension of these ideas is to study the effect of  excitations above the vacuum state or inclusion of new degrees of freedom in the form of flavours or defects.  Here it was understood that even  for  flavours or defects in the quenched approximation, the application of the Ryu-Takayanagi prescription \cite{Ryu:2006bv, Ryu:2006ef} appears to require knowledge of the backreaction from the corresponding probe degrees of freedom in the dual gravitational description  \cite{Chang:2013mca, Jensen:2013lxa, Kontoudi:2013rla, Jones:2015twa, Erdmenger:2015spo}. It has been subsequently pointed out in \cite{karchuhl} that this procedure can be circumvented by applying the gravitational entropy method of \cite{Lewkowycz:2013nqa} to the quenched degrees of freedom propagating in the un-backreacted gravitational backgrounds.

In this paper, we will study pointlike defects or ``impurities" that have a simple interpretation, namely they are test charges or heavy quarks introduced into the vacuum state of a large-$N$ QFT.  The coupling of the heavy quark to the quantum fields affects the entanglement of any region that contains the impurity, with the rest of the system. Specifically,  we  are interested in the {\em change} in entanglement entropy (EE)  of a spherical region of some radius $R$ upon introduction of a test quark in the ${\cal N}=4$ supersymmetric gauge theory in 3+1 dimensions, with $SU(N)$ gauge group.  This question becomes particularly interesting if one can deform the quantum mechanics of the pointlike impurity so that the system is not conformally invariant and the degree of entanglement is a nontrivial  function of the deformation strength.  Our goal will be to examine and identify general scale dependent properties of EE across different tractable examples of such impurities at strong 't Hooft coupling in the large-$N$ theory. 

In \cite{Lewkowycz:2013laa} the excess EE due to such heavy quarks in large rank symmetric and antisymmetric tensor representations were computed (both at weak and strong coupling) by exploiting conformal invariance and relating them to known results \cite{drukkerfiol, paper1, yamaguchi, passerini, paper2} for supersymmetric Wilson/Polyakov loops in the ${\cal N}=4$ theory.   In this paper we will apply the method of \cite{karchuhl} based on gravitational entropy contributions to obtain the EE excess due to the corresponding probes (D-branes) in the gravity dual, including the effect of deformations that trigger flows on the impurity.  The main results of this paper are summarized below:
\begin{itemize}
\item{ We focus attention on heavy quark probes in the symmetric and antisymmetric tensor representations of rank $k$, with $k\sim {\cal O}(N)$ (within the ${\cal N}=4$ theory at large-$N$), which are dual to D3 and D5-brane probes in AdS$_5\times$S$^5$. In the conformal case, the worldvolume of the probe contains an AdS$_2$ factor, reflecting the conformal nature of the quantum mechanics on the impurity. We calculate the contribution to the generalized gravitational entropy from these probe branes using the proposal of \cite{karchuhl} and find a match with the results of \cite{Lewkowycz:2013laa} deduced via independent arguments. A nontrivial aspect of the calculation and  observed agreement is the role played by 
the background Ramond-Ramond (RR) flux and its associated four-form potential, specifically in the case of the D3-brane probe dual to the symmetric representation source. The generalised gravitational entropy receives a contribution from the coupling of this potential to the D3-brane probe, and matching with the CFT arguments of \cite{Lewkowycz:2013laa} picks out a special choice of gauge for the four-form potential.
}
\item{We then study certain deformations on the probes which appear as  simple one-parameter BPS solutions for the D-brane embeddings. The D5-brane solution, first found in \cite{Callan:1998iq}, interpolates between $k$ sources in the fundamental representation at short distances, and an impurity transforming in the antisymmetric representation ${\cal A}_k$ at long distances.  The deformation appears as a dimensionful parameter $A$ in the UV\footnote{This is a puzzling aspect of both the D3- and D5-brane non-conformal solutions we study, as both appear to be triggered by the VEV of a dimension one operator in the UV picture \cite{Kumar:2016jxy}, and implies spontaneous breaking of conformal invariance, which should not be possible in  quantum mechanics (on the impurity).}, and has the effect of screening the fundamental sources into the representation ${\cal A}_k$.  This is most directly seen by examining the profiles of the gauge theory operators (e.g. ${\cal O}_{F^2}\,=\,{\rm Tr}F_{\mu\nu}F^{\mu \nu}\,+\,\ldots$) sourced by the impurity where the strength of the source first increases on short scales,  subsequently turns around and decreases monotonically (figure \ref{vev}) at large distances to an asymptotic value determined by the representation ${\cal A}_k$. 

We calculate the EE excess due to this impurity  within a spherical region of radius $R$ surrounding the source, by mapping the causal development of the spherical region to the Rindler wedge which is conformal to the hyperbolic space $H^3$ with temperature $\beta^{-1}=\frac{1}{2\pi}$.  The contribution of the probe to the gravitational entropy is obtained by varying the temperature of the dual hyperbolic AdS black hole. 
As a function of the dimensionless radius $(AR)$, we find that the EE excess displays the same qualitative behaviour 
(figure \ref{figd5flow}) as the profiles of gauge theory fields, namely an increase on short scales  accompanied by eventual decrease at large radii towards the asymptotic value governed by the representation ${\cal A}_k$.

 We also find that although the EE is a non-monotonic function of the radius $R$, the impurity free energy on $S^1\times H^3$ which can be interpreted as a relative entropy, increases monotonically from the UV to the IR.
  }
  \item{For the D3-brane probes, a simple BPS deformation exists which was discussed relatively recently in \cite{Schwarz:2014rxa} and \cite{Kumar:2016jxy}. There are two categories of these solutions (figure \ref{d3branefigs}): One yields a symmetric representation $({\cal S}_k)$ source in the UV ``dissociating" into $k$ coincident quarks in the IR, while the second category describes a heavy quark in representation ${\cal S}_k$ on the Coulomb branch of the ${\cal N}=4$ theory with $SU(N)$ broken to $U(1)\times SU(N-1)$. 
  \begin{figure}[h]
\begin{center}
\includegraphics[width=2.2in]{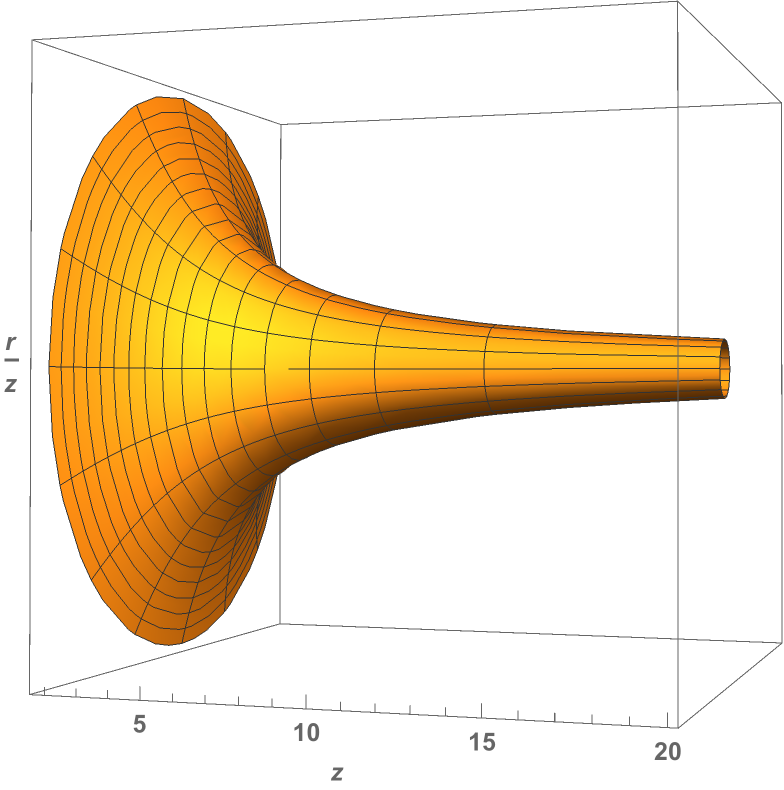}
\hspace{1.0in}
\includegraphics[width=2.2in]{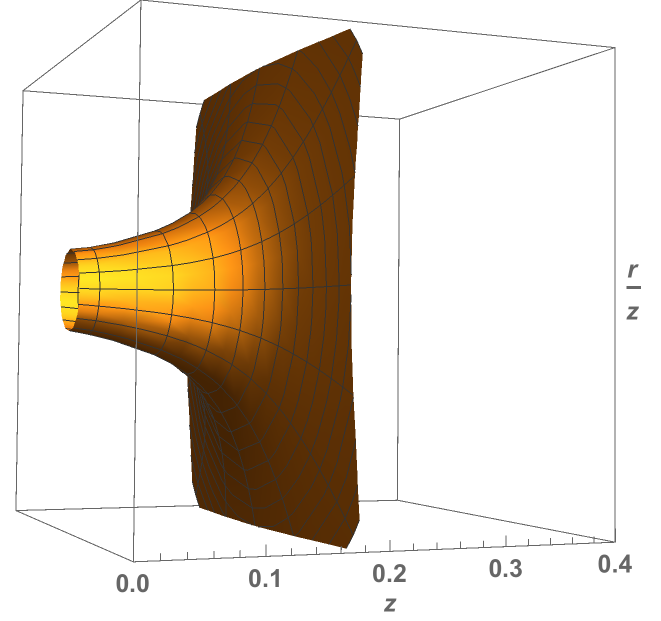}
\end{center}
\caption{\small{The two types of D3-brane embeddings in ${\rm AdS}_5$. Shown above are the proper sizes of the two-sphere wrapped by the D3-branes as a function of AdS radial coordinate $z$. The figure on the left represents an interpolation between the symmetric representation in the UV ($z\to 0$) and an IR spike of $k$ strings, while the one on the right is the symmetric representation source ending on a Coulomb branch D3-brane.
} }\label{d3branefigs}
\end{figure}
We apply the gravitational entropy method to these sources taking care to employ the correct gauge for the RR potential which yields the expected result for the undeformed conformal probe. In both cases the EE excess displays non-monotonic behaviour over short scales - first increasing as a function of $R$, and reaching a maximum. At large distances, however, the two categories display qualitatively distinct features. The EE excess for the first class of solutions saturates in the IR (figure \ref{d3flowaplus}) at a {\em higher} value (that of $k$ fundamental sources) than in the UV (corresponding to the representation ${\cal S}_k$). For the Coulomb branch solution, we find the the EE excess decreases monotonically in the IR without bound with some universal features (figure \ref{d3eeminus1}).

In all cases however, the free energy on $S^1\times H^3$ for each of the probes increases monotonically from the UV to the IR, consistent with the interpretation as a relative entropy. The IR asymptotics of this free energy for the Coulomb branch solution exhibits certain universal features, namely, quadratic and logarithmic dependence on the Coulomb branch VEV with the coefficient of the logarithmic term being universal.

We further confirm that D3-brane impurities with the deformations turned on,  display a screening of the source in the representation ${\cal S}_k$.  We see this for both categories of solutions by calculating the spatial dependence of gauge theory condensates sourced by the heavy quark impurities.
  }
  \end{itemize} 
  The paper is organized as follows: In section 2 we review the argument of \cite{karchuhl} for calculating the EE of probes without backreaction. We also  review known results for the EE of  conformal probes, and for completeness, we also explictly write out the trasnformations from AdS to AdS-Rindler and hyperbolic-AdS spacetimes. Section 3 is devoted to the analysis of the D5-brane probe embeddings and their entanglement entropies. In Section 4 we review the D3-brane BPS solutions. All details of the EE calculation for the D3-brane impurities are presented in Section 5. We summarize our results and further questions in Section 6.  Certain technical aspects of the calculations including transformations of D3-brane worldvolume integrals from one coordinate system to another and evaluation of certain integrals are relegated to the Appendix.

\section{Generalized gravitational entropy for probe branes}
It was argued in \cite{karchuhl} that the entanglement entropy contribution from a finite number $N_f$ of flavour degrees of freedom, introduced into a large-$N$ CFT (with a holographic gravity dual), can be computed without having to consider explicit backreaction of flavour fields. A key element in this approach is the method of \cite{Lewkowycz:2013nqa} which can, in principle, be adapted to include the backreaction from flavour fields. However, this turns out to be unnecessary as the leading contribution at order ${O}\left(N_f/N\right)$ is determined completely by an integral over the flavour branes in a geometry {\em without} backreaction.

The entanglement entropy of a spatial region $A_{d-1}$ in a CFT in $d$ spacetime dimensions can be calculated by a holographic version of the replica trick in Euclidean signature. This is performed by considering smooth, asymptotically AdS geometries with a finite size Euclidean circle at the conformal boundary of period $2\pi n$ (and $n\neq 1$) going around the boundary $\partial A_{d-1}$ of the spatial region of interest. The classical action for these geometries then yields the holographic entanglement entropy via,
\be
S(A_{d-1})\,=\,-\left.n\partial_n\left[\log Z(n)\,-\, n\log Z(1)\right]\right|_{n=1}\,.
\ee
This quantity only receives non-zero contribution from a boundary term within the bulk, arising from the locus of points where the circle shrinks.  This corresponds to  the Ryu-Takayanagi minimal surface \cite{Ryu:2006bv}. Upon introducing probe branes (defects or flavours) the complete action for the gravitational system can be separated into  `bulk' and `brane'  components:
\be
S_g\,=\,S_{\rm bulk}\,+\,\epsilon_0\,S_{\rm brane}\,,
\ee
where the brane contribution is parametrically smaller by a factor of $\epsilon_0\sim N_f/N$.  To paraphrase the argument of \cite{karchuhl}, if one views the backreacted metric as a small perturbation about (the $n$-fold cover of) AdS, the deviation of the bulk action from AdS only appears at order $\epsilon^2_0$. Then the probe contribution to the gravitational entropy at order $\epsilon_0$ is completely determined by an integral over the brane worldvolume alone. Furthermore, the brane embedding need only be known in ordinary AdS spacetime (with $n=1$), since the inclusion of backreaction will only affect the probe action at order $\epsilon^2_0$ and deviations of the embedding functions at order $(n-1)$ will also contribute to the action at order $(n-1)^2$, since the $n=1$ embedding solves the equations of motion.

To compute the entanglement entropy of the region $A_{ d-1}$ one applies the well known method of \cite{Casini:2011kv} for the specific case when $\partial A_{ d-1}$ is a sphere $S^{d-2}$. This maps the 
causal development of the region within the sphere to a Rindler wedge. The spherical boundary of the entangling region is mapped to the origin of the Rindler wedge. In this process the reduced density matrix for the degrees of freedom inside the sphere then corresponds to the Rindler thermal state with inverse temperature $\beta=2\pi$.
The latter is also conformal to a spacetime ${\cal H}$ with hyperbolic spatial slices $H^{d-1}$, so that ${\cal H}\simeq {\mathbb R}_t\times H^{d-1}$ \cite{Casini:2011kv}. The entanglement entropy of the region $A_{d-1}$ is then given by the thermal entropy of the CFT on ${\cal H}$:
\be
S\left(A_{d-1}\right)\,=\,\lim_{\beta \to 2\pi}\left(1\,-\,\beta\frac{\partial}{\partial\beta}\right)\log Z_{\cal H}\,.
%\qquad \log Z_{\}=-I_{grav.}
\label{tentropy}
\ee 
For theories possessing a holographic dual, the computation of $Z_{\cal H}$ requires a bulk (AdS) extension of the boundary Rindler wedge away from the Rindler temperature $\b=2\pi$.  This becomes possible for the case of a CFT where we may transform the bulk extension of the wedge to hyperbolically sliced ${\rm AdS}_{d+1}$ geometry.
The thermal partition function on ${\cal H}$ is computed holographically by the classical action of the bulk Euclidean ${\rm AdS}_{d+1}$ geometry with hyperbolic slices, the replica trick is implemented by allowing the inverse temperature of the hyperbolic black hole to deviate from the value $\beta=2\pi$:
\be
\log Z_{\cal H}\,=\,-I_{\rm AdS}(\beta)\,.
\ee
The unique extension of the bulk hyperbolically sliced geometry, away from $\b=2\pi$, is related to the replica method via the observation of \cite{Lewkowycz:2013nqa}. In particular, the value of the $n$ replicated partition  $Z(n)$ can be replaced by $n$ times the replicated partition function with the time interval restricted to the domain $[0, 2\pi)$, and eq.\eqref{tentropy} reduces to,
\be
S\left(A_{d-1}\right)\,=\,-\,\lim_{n\to1}\,n^2\pa_n\log Z_n\big|_{2\pi}\,,\qquad \log Z(n) \,=\,n\,\log Z_n\big|_{2\pi}\,.
\label{bentropy}
\ee
The method reproduces the vacuum EE area formula of \cite{Ryu:2006bv} and  will allow us to extract the EE excess due to the insertion of defects without the need for backreaction on either the background or the defect itself.  
\subsection{Conformal defects from D3/D5-branes and EE}
A point-like impurity in gauge theory arises most naturally upon the introduction of a Wilson line or heavy quark transforming in some representation of the gauge group.  Wilson lines in fundamental $(\Box)$, rank-$k$ antisymmetric $({\cal A}_k)$ and symmetric  $({\cal S}_k)$ representations of $SU(N)$ are particularly nice from the perspective of gauge/gravity duality as they have simple realisations in terms probe string and brane sources \cite{malpol,yamaguchi,paper1, paper2, passerini}.  Such sources compute BPS Wilson lines in different representations in the ${\cal N}=4$ supersymmetric gauge theory at strong coupling and large-$N$, and are introduced as probes in the dual ${\rm AdS}_5\times {\rm S}^5$ background. In the absence of any probe deformations, the world volume metric on such probes includes an ${\rm AdS}_2$ factor, so that the dual impurity theory is a (super)conformal quantum mechanics. 

The excess contribution from such an impurity to the EE of a spherical region in ${\cal N}=4$ SYM was calculated in \cite{Lewkowycz:2013laa} using the method described above, leading to eq.\eqref{tentropy} but where ${Z}_{\cal H}$ is replaced by the impurity partition function in hyperbolic space, computed by a Polyakov loop or circular Wilson loop $W_{\circ}$. One way to understand the appearance of the circular Wilson loop is to note that upon mapping  the causal development of a spherical region to the Rindler wedge, the worldline of the heavy quark maps to the  hyperbolic trajectory of a uniformly accelerated particle. Upon Euclidean continuation, the hyperbolic trajectory  turns into a circle.  Therefore, 
\be
S_{\rm imp}\,=\,\left(1-\beta\frac{\partial}{\partial\beta}\right)\,\ln W_{\circ}\left.\right|_{\beta=2\pi}\,=\,\ln W_\circ\left.\right|_{\beta=2\pi}\,+\,\,\int_{S^1_\beta\times{ H^3}}\sqrt g\,\langle T_{\tau \tau}\rangle_{W_{\circ}}\,,\label{EEcirc}
\ee
where in the final expression we are required to compute the expectation value of the field theory stress tensor on ${\cal H}$, in the presence of the Wilson/Polyakov loop insertion. As argued in \cite{Lewkowycz:2013laa}, conformal invariance fixes the form of the stress tensor, and the expectation value of the energy density integrated over $S^1_\beta\times {H}^3$ depends on a single normalisation constant $h_w$:
\be
\int_{S^1_\beta\times{{H}^3}}\sqrt g\,\langle T_{\tau\tau}\rangle_{W_{\circ}}\,=\,-8\pi^2 h_w\,.
\ee 
The normalisation constant $h_w$ for ${\cal N}=4$ SYM was calculated in \cite{Gomis:2008qa} by relating it to the expectation of a dimension two chiral primary field, with net result,
\be
S_{\rm imp}\,=\,\left(1\,-\,\frac{4}{3}\lambda\partial_\lambda\right)\,\ln W_\circ\,.\label{EEcirc}
\ee
While localization results can, in principle, be used to determine the circular Wilson loop in various representations for any $N$ and gauge coupling, we will focus attention on the strict large-$N$ limit at strong 't Hooft coupling $\lambda \to \infty$ \cite{paper1, drukkerfiol}.
In this limit, the following results can be deduced for the EE contributions from the conformal impurities in the three different representations described above\footnote{The results quoted here differ from those of \cite{Lewkowycz:2013laa} by an overall factor of $1/2$. We clarify the reason for this normalization below eqs.\eqref{rindlerads} and \eqref{f1check}. }:
\bea
&& S_\Box\,=\,\frac{\sqrt{\lambda}}{6}\,,\label{EElist}\\\nonumber\\\nonumber
&&S_{{\cal A}_k}\,=\,\frac{ N}{9\pi}\sqrt{\lambda}\,\sin^3\theta_k\,,\qquad
\pi(1-\kappa)\,=\,\theta_k\,-\,\sin\theta_k\,\cos\theta_k\,,\qquad\kappa\,\equiv\,\tfrac{k}{N}\,,\\\nonumber\\\nonumber
&& S_{{\cal S}_k}\,=\, N \left(\sinh^{-1}\tilde\kappa\,-\,\tfrac{1}{3}\tilde\kappa\sqrt{\tilde\kappa^2+1}\right)\,,\qquad \tilde\kappa\,\equiv\,\tfrac{\sqrt{\lambda}\,k}{4N}\,.
\eea
Our aim will be to reproduce these results for the conformal impurities using the method of \cite{karchuhl} and then apply the same to the case of the non-conformal impurity flows that were discussed in \cite{Kumar:2016jxy}.  

\subsection{From AdS to hyperbolic AdS}

Now we review the maps that take the AdS-extension of the causal development of the  spatial sphere in ${\mathbb R}^{1,3}$ to hyperbolically sliced ${\rm AdS}_5$. This will help set our conventions, and will be important  subsequently since the evaluation of EE for  non-conformal impurities will involve computation of integrals over specific brane embeddings in hyperbolic-AdS geometry, and the explicit calculation of these will require us to go back and forth between different coordinate systems.

We first consider the transformation,
\bea
&& x_{\alpha}\,=\,\frac{\tilde x_{\alpha}\,+\,\frac{c_{\alpha}}{2R}\,\left(\tilde x^2+\tilde z^2\right)}{1\,+\,\frac{c}{R}\cdot\tilde x\,+\,\frac{c^2}{4R^2}\,\left(\tilde x^2+z^2\right)}\,-\,c_{\alpha}\,R\,,\qquad \alpha\,=\,0,\ldots3\,,
\label{specialconformal}\\\nonumber\\\nonumber
&& z\,=\,\frac{\tilde z}{1\,+\, \frac{c}{R}\cdot\tilde x\,+\,\frac{c^2}{4R^2}\,\left(\tilde x^2\,+\,z^2\right)}\,,
\eea
where $c^{(\alpha)}\,=\,(0,1,0,0)$. Here $z$ is the radial AdS coordinate, with the conformal boundary at $z=0$. This is the extension of the boundary CFT special conformal transformation to an isometry of AdS$_5$. The map has the following actions:

\begin{itemize}
\item{
{\em On the conformal boundary} at $z=0$,  the ball ${\cal B}$: $x_{1}^2\,+\,x_{2}^2\,+\,x_{3}^2 \,\leq R^2$ at $x_0=0$ is mapped to the half-line $\tilde x_{1} \geq 0$. The causal development of ${\cal B}$ is mapped to the Rindler wedge $\tilde x_1 > |\tilde x_0|$.}
\item{The world line of the impurity on the boundary, located at the spatial origin $x_i=0$, is mapped to the trajectory of a uniformly accelerated particle,  $\tilde x_1^2\,-\, \tilde x_0^2 \,=\,4R^2$, with $\tilde x_1 >0$. In Euclidean signature this maps to {\em one half} of the circular Wilson loop with $\tilde x_1 >0$.}
\item{The transformation acts on the ${\rm AdS}_5$ Poincar\'e  patch metric as an isometry:
\be
ds^2\,=\,\frac{dz^2\,+\,dx_\alpha dx^\alpha}{z^2}\quad\to\quad\frac{d\tilde z^2\,+\,d\tilde x_\alpha d\tilde x^\alpha}{\tilde z^2}\,,
\ee
while the boundary metric itself transforms by a conformal factor. The holographic extension of the causal development of the ball ${\cal B}$ into the AdS bulk (entanglement wedge) is given by the causal development of the hemisphere $z^2\,+\,x_1^2\,+\,x_2^2\,+\,x_3^2\,=\,R^2$ (defined at $x_0=0$). This is mapped by the above isometry to the Rindler-AdS wedge $\tilde x_1 \geq \tilde x_0$.
}
\end{itemize}
The Rindler-AdS wedge is further mapped to hyperbolically sliced AdS$_5$ by the transformations listed below. First we parametrize the Rindler-AdS wedge by defining the coordinates,
\be
\tilde x_1\,=\, r_1 \cosh t\,,\qquad \tilde x_0\,=\,r_1\,\sinh t\,,\qquad \tilde x_2\,=\,r_2\cos\phi\,,\qquad \tilde x_3\,=\,r_2 \sin\phi\,,
\ee
so that 
\be
ds^2\big|_{\rm Rindler-AdS}\,=\,\frac{1}{\tilde z^2}\left(d\tilde z^2\,+\,dr_1^2\,-\,r_1^2dt^2\,+\, dr_2^2\,+\,r_2^2d\phi^2\right)\,.
\ee
The wordline of the heavy quark on the boundary is given by $r_1\,=\,2 R$.
In order to perform the replica trick it is crucial that we move to Euclidean signature, via the replacement $t\to i\tau$, so we obtain AdS in ``double polar" coordinates, and the heavy quark impurity then traces out a Polyakov loop at $r_1= 2R$,
\bea
 ds^2_{\rm E}\,=\,\frac{1}{\tilde z^2}\left(d\tilde z^2\,+\,dr_1^2\,+\,r_1^2d\tau^2\,+\, dr_2^2\,+\,r_2^2\,d\phi^2\right)\,,\qquad -\frac{\pi}{2} \leq \tau \leq \frac{\pi}{2}\,.\label{rindlerads}
\eea
The Euclidean time $\tau$ must be restricted to the domain where $\cos \tau$ is positive, so that $\tilde x_1 >0$. The $\tau$-coordinate is periodic under the shifts $\tau \to \tau +2\pi$ which also ensures that the ``double polar" geometry is free of conical singularities.
The map to hyperbolically sliced AdS$_5$ is achieved by the transformations
\bea
&&\tilde z\,=\,\frac{2R}{\rho\,\omega}\,,\qquad r_1\,=\,\frac{2R}{\rho\, \omega}\,\sqrt{\rho^2 \,-\,1}\,,\qquad r_2\,=\, \frac{2 R}{\omega}\sinh u\,\sin\theta\,,
\label{rindlertohyp}\\\nonumber\\\nonumber
&&\omega\,=\,\left(\cosh u\,-\,\sinh u\cos\theta\right)\,,
\eea
which yield the Euclidean AdS$_5$ black hole with hyperbolic horizon,
\be
ds^2\big|_{\rm AdS-Hyp}\,=\,\frac{d\rho^2}{\rho^2\,-\,1}\,+\,(\rho^2-1)\,d\tau^2\,+\,\rho^2 \left(du^2\,+\,\sinh^2u\,d\Omega_2^2\right)\,.\label{hypads}
\ee
Once again we have the restriction $-\frac{\pi}{2}\leq \tau \leq \frac{\pi}{2}$  on the range of the Euclidean time which has periodicity $2\pi$, guaranteeing that the space caps off smoothly at $\rho=1$. Finally, it will be useful to to recall the coordinate transformations which directly map the entanglement wedge in the original AdS spacetime,
\be
ds^2\big|_{\rm AdS}\,=\,\frac{1}{z^2}\left(dz^2\,-dx_0^2\,+\,dr^2\,+\,r^2\,d\Omega_2^2\right)\,,
\ee
to the hyperbolic AdS$_5$ black hole \eqref{hypads} with inverse temperature $2\pi$. The relevant coordinate transformations are (in Lorentzian signature):
\bea
z\,=\,\frac{R}{\rho\cosh u\,+\,\sqrt{\rho^2-1}\cosh t}\,,\qquad
x_0\,=\,\sqrt{\rho^2-1}\,z\,\sinh t\,,\qquad r\,=\,\rho\,z\,\sinh u\,.\nonumber
\\ \label{poinctohyp}
\eea
Upon continuation to imaginary time $t = i\tau$, we must restrict to the domain of $\tau$ to $-\frac{\pi}{2}\leq \tau \leq \frac{\pi}{2}$. It can be shown that the pre-image of the Euclidean hyperbolic AdS black hole, given this domain, is the interior of the hemisphere in the original (Euclidean) AdS geometry,
\be
x_0^2\,+\,r^2\,+\,z^2\,\leq\,R^2\,,\qquad r, z \geq 0\,.
\ee 

\paragraph{Hyperbolic AdS and replica method:} The replica method requires that we consider a hyperbolic AdS black hole in which the Euclidean time has  period $2\pi n$ where $n\neq 1$, so that
\bea
&&ds^2\big|_{\rm AdS-Hyp}\,=\,\frac{d\rho^2}{f_n(\rho)}\,+\,
f_n(\rho)\,d\tau^2\,+\,\rho^2\left(du^2\,+\,\sinh^2u\,d\Omega_2^2\right)\,,
\label{hypAdSmetric}
\\\nonumber\\\nonumber
&& f_n(\rho)\,=\,\rho^2\,-\,1\,-\frac{\rho_+(\rho_+^2-1)}{\rho^2}\,,\qquad\qquad
\rho_+\,=\,\frac{1}{4n}\left(1\,+\,\sqrt{1+8n^2}\right)\,.
\eea
The Hawking temperature of the black hole is
\be
\beta^{-1}\,=\,T_H\,=\,\frac{f'(\rho_+)}{4\pi}\,=\,\frac{2\rho_+^2 - 1}{2\pi \rho_+}\,.
\label{Thawk}
\ee
It is clear that implementation of the replica trick is equivalent to varying the Hawking temperature of the black hole, ensuring as usual, the absence of a conical singularity in the Euclidean geometry. In this approach the entanglement entropy is given by the thermal entropy evaluated in the hyperbolic AdS geometry. In particular, using eq.\eqref{bentropy}, we have
\be
S\,=\,\lim_{\beta\to 2\pi}\,\beta\,\partial_\beta {\rm I}_{2\pi}(\beta)\,.\label{hypthermalEE}
\ee
Here ${\rm I}_{2\pi}(\beta)$ is the action of the hyperbolic AdS geometry including any probes dual to the impurities or defects under consideration, and where the integration over Euclidean time is restricted to the domain $\left[0, 2\pi \right)$.
\subsection{Warmup: A single fundamental quark}
As a warmup, we compute the EE excess due to the insertion of a single fundamental quark into the spherical entangling region. In the AdS dual, this is achieved by inserting a probe fundamental string (F1) into the hyperbolic AdS geometry and computing the thermal entropy from the Nambu-Goto worldsheet action in this geometry. The F1-string worldsheet is placed at $u=0$, and  stretches from the hyperbolic horizon at $\rho=\rho_+$ to the conformal boundary at $\rho \,=\, \rho_\infty \to \infty$. The tension for the fundamental string, in units of the AdS radius is
\be
T_{\rm F1}\,=\,\frac{1}{2\pi \alpha'}\,=\,\frac{\sqrt{\lambda}}{2\pi}\,,
\ee
where $\lambda$ is the 't Hooft coupling for the ${\cal N}=4$ theory. Then the action for the static F-string embedding stretched along the radial AdS coordinate is
\be
{\rm I}_{\rm F1}(\beta)\,=\,\frac{\sqrt{\lambda}}{2\pi}\,\int_{\rho_+}^{\rho_\infty} d\rho\int_{-\frac{\pi}{2}}^{\frac{\pi}{2}} d\tau\,\sqrt{\det *g}\,+\,{\rm I}_{\rm F1\,c.t}\,.
\ee
The determinant of the induced metric $*g$ on the worldsheet for this embedding is unity, and the boundary counterterm $I_{\rm F1\,c.t.}$ which regularises the worldsheet action is independent of the temperature $\beta$ as it is only sensitive to UV details.
Varying with respect to $\beta$, we thus obtain
\be
S_{\Box}\,=\,\beta\,\frac{\partial\, {\rm I}_{\rm F1}(\beta)}{\partial\beta} \,=\,\frac{\sqrt \lambda}{6}\,. \label{f1check}
\ee
Our result differs by a factor of two from that of \cite{Lewkowycz:2013laa}, as the range of integration over Euclidean time is restricted to $-\frac{\pi}{2}\leq \tau \leq \frac{\pi}{2}$, which corresponds to one half of the Polyakov loop on $S^1_{\beta}\times { H}^3$.
\paragraph{EE from stress tensor evaluation:}
For this simple example it is instructive to verify how the above result can be reproduced holographically, using eq.\eqref{EEcirc} which relies on the expectation value of the stress tensor in the presence of the temporal Wilson line in  Rindler frame. This computes the expectation value of the entanglement Hamiltonian which generates time translations along the compact time direction. In particular, the EE for the impurity is given as
\be
S_{\,\Box}\,=\,\ln Z_{\cal H}^{\,\Box}\,+\,\int_{\cal H}\,\sqrt{g_{\cal H}}\,\langle T_{\tau\tau}\rangle_{\Box}\,.
\ee
The ingredients in the computation can be calculated either directly  in the AdS Poincar\'e patch, or after translating to the hyperbolic AdS picture. In the Poincar\'e patch, we need to ensure that all integrals over the Euclidean string worldsheet are restricted to the domain,
\be
{\cal D}:\quad x_0^2\,+\,z^2\,\leq R^2\,,  \qquad z > 0\,.\label{domain}
\ee
Therefore, the impurity action in hyperbolic space is given by integrating the  (Euclidean) Nambu-Goto action in the Poincar\'e patch of AdS over ${\cal D}$:
\be
-\ln Z_{\cal H}^{\,\Box}\,=\,{\rm I}_{\Box}\,=\,\frac{\sqrt \lambda}{2\pi}\left[\int_{\epsilon}^R dz \,\frac{1}{z^2}\int_{-\sqrt{R^2-z^2}}^{\sqrt{R^2-z^2}} dx_0\,-\,\int_{-R}^R dx_0 \,\frac{1}{\epsilon}\right]\,=\,-\frac{\sqrt\lambda}{2}\,.
\ee
The second term is the worldsheet counterterm induced on the conformal boundary at $z=\epsilon$, as $\epsilon$ is taken to zero. The stress tensor expectation value{\footnote{The worldsheet stress tensor for the string embedding is obtained by varying with respect to the spacetime metric, so that $T_{\alpha\beta}\,=\,-2\frac{\partial{\cal L}}{\partial g^{\alpha \beta}}$, in Lorentzian signature.}} for the heavy quark source in the Unruh state, or equivalently, in hyperbolic space ${\cal H}$ would normally be computed by reading off the normalizable mode of the metric sourced by the probe string in the bulk.  Alternatively, from the Hamiltonian formulation of the AdS/CFT correspondence, the (regularized) energy of the probe should directly yield the energy of the corresponding source (impurity) in the boundary CFT \cite{Karch:2008uy}.  The result for the energy of the probe string is thus of the form
\be
\int_{\cal H}\langle T_{\tau\tau}\rangle_{\Box}\,=\,\frac{\sqrt{\lambda}}{2\pi}\left[\int_{-\pi/2}^{\pi/2} d\tau\int_1^{\rho_\infty} d\rho\,g_{\tau\tau}\right]\,,
\ee
where $\rho_\infty$ is the UV cutoff. Keeping only the finite terms, we find
\be
\int_{\cal H}\langle T_{\tau\tau}\rangle_{\Box}\,=\,-\, \frac{\sqrt\lambda}{3}\,,
\ee
so that the contribution to the EE of the spherical region from the heavy quark is 
\be
S_{\,\Box}\,=\,\frac{\sqrt{\lambda}}{6}\,.
\ee
\section{D5-brane impurity}
In this section we will focus our attention on the D5-brane embedding which computes the BPS Wilson loop in ${\cal N}=4$ SYM, in the antisymmetric tensor representation.  The embedding admits a deformation which can be interpreted as an RG flow on the worldvolume of the  impurity \cite{Kumar:2016jxy}. Our goal will be to extract the behaviour of the impurity EE along this flow.
\subsection{ AdS embeddings of the D5-brane}
The D5-brane embedding, dual to a straight Wilson line in the ${\cal N}=4$ theory, preserves an $SO(5)$ subgroup of the global R-symmetry.  This is realized geometrically, by having the D5-brane wrapping an $S^4$ latitude of the five-sphere in AdS$_5 \times{\rm S}^5$.   In the  non-conformal ``flow'' solution described in \cite{Kumar:2016jxy}, the polar angle $\theta$ associated to this $S^4$ latitude varies as a function of the radial position in AdS$_5$.  We can choose the worldvolume coordinates to be 
$(\sigma, x_0, \Omega_4)$, where $\sigma$ parametrises the non-compact spatial coordinate on the brane. We will eventually choose the gauge $\sigma=z$.
The induced metric for such an embedding in (Euclidean) AdS$_5 \times {\rm S}^5$ is, 
\be
*ds^2\,=\,d\sigma^2\left(\frac{z'(\sigma)^2}{z^2}\,+\,\theta'(\sigma)^2\right)\,+\,\frac{dx_0^2}{z^2}\,+\, \sin^2\theta\,d\Omega_4^2\,.
\ee 
The action for the D5-brane consists of the standard Dirac-Born-Infeld (DBI) and Wess-Zumino (WZ) terms. The latter supports the configuration when a non-zero, radial world-volume electric field $F_{0z}$ is switched on.  In Euclidean signature this is purely imaginary and will be denoted in terms of the real quantity $G$:  
\be
G\,=\,-\,2\pi i\alpha^\prime\,F_{0\sigma}\,.
\ee
The Wess-Zumino term for the D5-brane embedding is induced by the pullback of the RR four-form potential $C_{(4)}$ determined by the volume form on AdS$_5 \times {\rm S}^5$. In particular, the relevant component of $C_{(4)}$ is
\be
C_{(4)}\,=\,\frac{1}{g_s}\left[\frac{3}{2}\left(\theta-\pi\right)\,-\,
\sin^3\theta\,\cos\theta\,-\,\frac{3}{2}\sin\theta\cos\theta
\right]\,\omega_4\,,
\ee
where $\omega_4$ is the volume form of the unit four-sphere. The four-form potential is chosen so that the five-form flux comes out proportional to the volume form of ${\rm S}^5$:
\be
F_{(5)}\,=\,dC_{(4)}\,=\,\frac{1}{g_s}\,4\sin^4\theta\,d\theta\wedge \omega_4\,.
\ee
The D5-brane embedding is then determined by the equations of motion following from the action
\be
{\rm I}_{\rm D5}\,=\,{\rm T}_{\rm D5}\,\int d^6\sigma\,e^{-\phi}\sqrt{*g\,+\,2\pi\alpha^\prime\,F}\,-i\, g_s\,{\rm T}_{\rm D5}\,\int 2\pi\alpha^\prime F\wedge C_{(4)}\,+\,{\rm I}_{\rm c.t.}\,.
\ee
The action is regularized by counterterms ${\rm I}_{\rm c.t.}$. The dilaton $\phi$ vanishes in the AdS$_5\times{\rm S}^5$ background dual to the ${\cal N}=4$ theory, and the D5-brane tension can be expressed in terms of gauge theory parameters as 
\be
{\rm T}_{\rm D5}\,=\,\frac{N\sqrt\lambda}{8\pi^4}\,,\qquad \lambda\,=\, 4\pi g_s N\,.
\ee
The counterterms can be split in two pieces: one which regulates the UV divergences in the action and another which fixes the number of units of string charge carried by the embedding to be $k\in \mathbb Z$ \cite{drukkerfiol, Kumar:2016jxy},
\bea
&&{\rm I}_{\rm c.t.}\,=\,{\rm I}_{\rm UV}\,+\,{\rm I}_{U(1)} \,,\\\nonumber
\\\nonumber
&& {\rm I}_{\rm UV}\,=\,-\,\int dx_0\,\left(z\,\frac{\delta \,{\rm I}}{\delta \left(\partial_\sigma z\right)}\,+\,\left(\theta(\sigma)\,-\,\theta\big|_{z=0})\right)\,\frac{\delta\, {\rm I}}{\delta \left(\partial_\sigma \theta\right)}\right)\Big|_{z\,=\,\epsilon}\,.\\\nonumber\\\nonumber
&&{\rm I}_{\rm U(1)}\,=\,-i\int dx_0 \,d\sigma\,F_{\mu\nu}\,\frac{\delta {\rm I}}{\delta F_{\rm \mu\nu}}\,=\,ik\int dx_0\,d\sigma\,F_{0\sigma}\,.
\eea
The counterterm ${\rm I}_{\rm U(1)}$ enforces a Lagrange multiplier constraint that fixes the number of units of string charge carried by the configuration.
  Putting together all these ingredients, choosing the gauge $\sigma\,=\,z$, the final form for the D5-brane action  is
\be
{\rm I}_{\rm D5}\,=\,{\rm T}_{\rm D5}\frac{8\pi^2}{3} \int dx_0\int_\epsilon dz\left[\sin^4\theta\sqrt{z^{-4}\,+\,z^{-2}\,\theta^{\prime 2}\,-\,G^2}\,-D(\theta)\, G\right]\,+\,{\rm I}_{\rm UV}\,,
\ee
with 
\be
D(\theta)\,\equiv\,\sin^3\theta\cos\theta\,+\,\frac{3}{2}\,\left(\sin\theta\cos\theta\,-\,\theta\,+\,\pi(1-\kappa)\right)\,,\qquad\kappa\,\equiv\,\frac{k}{N}\,.
\ee
\subsubsection{The constant embedding}
It is easy to check that the equations of motion yield a {\em constant} solution:
\be
\theta\,=\,\theta_{\kappa}\,,\qquad \sin\theta_{\kappa}\cos\theta_\kappa\,-\,\theta_\kappa\,+\,\pi(1-\kappa)\,=\,0\,.
\ee
This solution is BPS and has vanishing regularized action in Poincar\'e patch. It yields the straight BPS Wilson loop in the antisymmetric tensor representation ${\cal A}_k$ \cite{yamaguchi, passerini, paper1}.  In all respects the constant solution is identical to the F-string solution for a fundamental quark, except for the normalization of the action which is controlled by $\theta_\kappa$.

\paragraph{Embedding in hyperbolic AdS:} The contribution to the EE of a spherical region can be calculated by applying the formula eq.\eqref{hypthermalEE} to the constant embedding in hyperbolic AdS space \eqref{hypAdSmetric}.  Repeating the above excercise for the solution which yields $\theta\, =\, \theta_\kappa$, we obtain the regularized action as a function of the temperature of the hyperbolic AdS black hole:
\be
{\rm I}_{\rm D5}(\beta)\,=\, {\rm T}_{\rm D5 }\frac{8\pi^2}{3}\int_{-\frac{\pi}{2}}^{\frac{\pi}{2}} d\tau\int_{\rho_+}^{\rho_\infty}d\rho\left[\sin^4\theta_\kappa\sqrt{1\,-\,G^2}\,-\,D(\theta_\kappa)\,G\right]\,+\,{\rm I}_{\rm UV}\,.
\ee
where $G=-\cos\theta_\kappa$.
The entanglement entropy contribution from the impurity in the antisymmetric tensor representation is then,
\be
S_{{\cal A}_k}\,=\,\lim_{\beta\to 2\pi}\beta\partial_\beta{\rm I}_{\rm D5}(\beta)\,=\,\frac{N}{9\pi}\,\sqrt{\lambda}\,\sin^3\theta_\kappa\,.
\ee
\subsubsection{The D5 flow solution}
The Poincar\'e patch action for the D5-brane embedding permits a non-constant zero temperature BPS solution \cite{Callan:1998iq}. This solution interpolates between a spike or bundle of $k$ coincident strings in the UV and the blown-up D5-brane configuration corresponding to the antisymmetric representation ${\cal A}_k$ reviewed above. In the boundary gauge theory, the flow can be interpreted as the screening of  $k$ coincident quarks in the fundamental representation to a source in the antisymmetric tensor representation \cite{Kumar:2016jxy}.  As seen in \cite{Kumar:2016jxy}, the flow appears as a result of a condensate for a dimension one operator in the UV worldline quantum mechanics of the impurity.  The Poincar\'e patch BPS embedding solves the first order equation,
\be
z\,\frac{d\theta}{dz}\,=\,-\,\frac{\partial_\theta \tilde D}{\tilde D}\,,\qquad \tilde D(\theta)\,\equiv\,\left(\sin^5\theta\,+\,D(\theta)\,\cos\theta\right)\,,
\label{bps}
\ee
and is explicitly given by the solution,
\be
\frac{1}{z}\,=\,\frac{A}{\sin\theta}\left(\frac{\theta\,-\,\sin\theta\cos\theta\,-\,\pi(1-\kappa)}{\pi\kappa}\right)^{1/3}\,,
\ee
where $A$ is an integration constant with dimensions of inverse length.
For small $z$,  the polar angle $\theta$ approaches $\pi$, so that the $S^4$ wrapped by the D5-brane shrinks to zero size and the collapsed configuration must be viewed as $k$-coincident strings. In the IR limit on the other hand, when $z \gg 1/A$, $\theta$ approaches $\theta_\kappa$ which yields the blown-up D5-brane embedding. 

In order to calculate the excess EE contribution from this non-conformal impurity in the boundary CFT, we first need to map the configuration to hyperbolically sliced AdS \eqref{hypAdSmetric}. The internal angle $\theta$ of the ten dimensional geometry is unaffected by the map. The only other active coordinate in the D5-brane embedding is the radial position $z$  in AdS spacetime which, upon rewriting in terms of hyperbolic Euclidean AdS coordinates \eqref{poinctohyp}, yields the transformed solution:
\be
\frac{1}{R}\left(\rho\,+\,\sqrt{\rho^2-1}\,\cos\tau\right)\,=\,\frac{A}{\sin\theta}\left(\frac{\theta\,-\,\sin\theta\cos\theta\,-\,\pi(1-\kappa)}{\pi\kappa}\right)^{1/3}\,,
\ee
with the restriction $-\frac{\pi}{2} \leq \tau \leq \frac{\pi}{2}$. 
The impurity is placed at the spatial origin, $r=0$ in ${\mathbb R}^4$, which corresponds to $u=0$ in $S^1\times{H}^3$. Since $\theta$ is a function of $\rho$ and $\tau$, the induced metric on the D5-brane is,
\bea
*ds^2\big |_{\rm D5}&&=\,\left[f_n(\rho)\,+\,\left(\partial_\tau\theta\right)^2\right]\,d\tau^2\,+\,
\left[\frac{1}{f_n(\rho)}\,+\,\left(\partial_\rho\theta\right)^2\right]\,d\rho^2\,+\,2\partial_\rho\theta\, \partial_\tau\theta\,d\tau\,d\rho\,\nonumber\\
&& +\,\sin^2\theta\,d\Omega_4^2\,,
\eea
where $f_n(\rho)$ is given in eq.\eqref{hypAdSmetric}.
The D5-brane embedding, mapped to hyperbolic AdS, must also have a non-trivial background worldvolume electric field. Since the embedding shares only the temporal and radial directions with the bulk AdS$_5$ geometry, there is only one component of the field strength to switch on:
\be
i\tilde G\,=\,2\pi \alpha^\prime\,F_{\tau\rho}\,.
\ee
For the case with $\beta=2\pi$, $F_{\tau\rho}$ can be obtained directly by transforming the field strength in the Poincar\'e patch solution. To implement the replica trick, however, we first need to consider general temperatures of the hyperbolic black hole.  Using the above ansatz for the D5-brane embedding, the action in the hyperbolic AdS background is,
\bea
{\rm I}_{\rm D5}(\beta)\,=&&{\rm T}_{\rm D5}\,{\rm Vol}({\rm S}^4)\int_{-\frac{\pi}{2}}^{\frac{\pi}{2}} d\tau\int_{\rho_+}^{\rho_\infty} d\rho\left[\sin^4\theta\sqrt{1\,-\,\tilde G^2\,+\,f_{n}(\rho)\,\left(\partial_\rho\theta\right)^2\,+\,\frac{(\partial_\tau\theta)^2}{f_n(\rho)}}\right.\nonumber\\\nonumber\\
&&\left.-\,D(\theta) \tilde G\right]\,+\,{\rm I}_{\rm UV}\,.
\eea
Solving for $\tilde G$ using its equation of motion and plugging it back in, 
\bea
{\rm I}_{\rm D5}(\beta)\,=\,{\rm T}_{\rm D5}\frac{8\pi^2}{3}\int_{-\frac{\pi}{2}}^{\frac{\pi}{2}} d\tau\int_{\rho_+}^{\rho_\infty} d\rho &&\sqrt{\sin^8\theta\,+\,D(\theta)^2}\,\times\label{offshellac}\\\nonumber\\\nonumber
&&\sqrt{1\,+\,f_{n}(\rho)\,\left(\partial_\rho\theta\right)^2\,+\,\frac{(\partial_\tau\theta)^2}{f_n(\rho)}}\,+\,{\rm I}_{\rm UV}\,.
\eea
In order to extract entanglement entropy excess due to the impurity, we need to vary this action with repect to $\beta$ and set $\beta=2\pi$, whilst keeping fixed $\theta(\rho,\tau)$ as the BPS solution at $\beta=2\pi$. The latter is justified  because the first variation of the action with repect to $\theta$ vanishes by the equations of motion at $\beta=2\pi$.   

Once the variations with respect to $\beta$ are performed, the remaining integrals are most easily evaluated in Poincar\'e patch coordinates, in which the D-brane embedding function is simpler.  The transformations \eqref{poinctohyp} when restricted to the location of the heavy quark at $u=r=0$ imply,
\be
\rho\,=\,\frac{R^2 + x_0^2 +z^2}{2zR}\,,\qquad \cos\tau\,=\,\frac{R^2-x_0^2-z^2}{\sqrt{(x_0^2+z^2+R^2)^2\,-\, 4R^2z^2}}\,.
\ee  
The Jacobian for the transformation on the worldvolume back to Poincar\'e patch coordinates is, 
\be
\left|\frac{\partial \rho}{\partial z}\frac{\partial \tau}{\partial x_0}\,-\,\frac{\partial \rho}{\partial x_0}\frac{\partial \tau}{\partial z}\right|\,=\,\frac{1}{z^2}\,.
\ee
We also note that the kinetic terms for a static Poincar\'e patch configuration satisfy,
\be
z^2 \theta'(z)^2\,=\,(\rho^2-1)(\partial_\rho \theta)^2\,+\,\frac{(\partial_\tau \theta)^2}{\rho^2-1}\,.
\ee
We first evaluate the action (or free energy) of the BPS embedding in the hyperbolic AdS background with $\beta=2\pi$, by recasting in Poincar\'e patch coordinates:
\bea
{\rm I}_{\rm D5}(2\pi)\,=\,{\rm T}_{\rm D5}\,{\rm Vol({\rm S}^4)}\int\int_{\cal D}dx_0\,dz\,\frac{d}{dz}&&\left[\frac{-1}{z}\tilde D(\theta)\right]\,-\,\frac{2R}{\epsilon}\tilde D(\theta)\big |_{z=\epsilon}\,,
\eea
where $\tilde D(\theta)$ is defined in eq.\eqref{bps}. 
Although the integrand is a total derivative, the  fact that the integration region is limited to the half-disk ${\cal D}$ (eq.\eqref{domain}), renders the evaluation  nontrivial. In particular, the integration over $x_0$ is performed first since the integrand is independent of time. Following this, the remaining integral can be performed numerically after exchanging the integration variable $z$ for $\theta$, which is more convenient as the solution is known explicitly for $z$ as a function of $\theta$.
\begin{figure}[h]
\begin{center}
\includegraphics[width=3in]{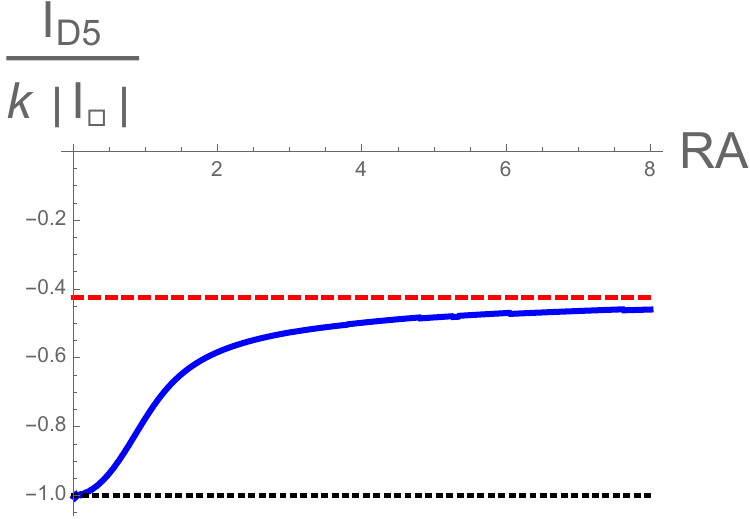}
\end{center}
\caption{\small{The free energy $-\ln Z_{\cal H}\,=\,  {\rm I}_{\rm D5}$ of the non-conformal D5-brane impurity (blue) on $H^3$ with $\beta=2\pi$, $\kappa=0.5$ as a function of the deformation parameter $A$. It interpolates between the values of the circular Wilson lines for $k$
coincident quarks (dotted black) in the fundamental representation, and the antisymmetric tensor representation ${\cal A}_k$ (dashed red line).}}\label{actionplot}
\end{figure}
The values of the (regularized) actions for the two types of conformal sources, fundamental and antisymmetric tensor ${\cal A}_k$ in hyperbolic space are:
\be
k\,{\rm I}_{\,\Box}(2\pi)\,=\,-\,k\frac{\sqrt\lambda}{2}\,,\qquad\qquad
{\rm I}_{{\cal A}_k}(2\pi)\,=\,-\,\frac{N\sqrt\lambda}{3\pi}\,\sin^3\theta_{\kappa}\,.
\ee
The partition function $\ln Z_{\cal H}$ of the heavy quark source in  hyperbolic space with inverse temperature $\beta=2\pi$ is plotted in figure \ref{actionplot} as a function of the deformation parameter $A$.  It is a monotonically decreasing function of the size of the entangling region and interpolates between the value for $k$ coincident fundamental quarks in the UV and that for a source transforming in the antisymmetric tensor representation ${\cal A}_k$ in the IR.

We note that $-\ln Z_{\cal H}$ is like a relative entropy \cite{relative}. It is the free energy difference between the embeddings with non-zero and vanishing deformations $A$ in the thermal state with $\beta=2\pi$ associated to the modular Hamiltonian. This explains the monotonic increase of $-\ln Z_{\cal H}$ with $AR$, and the vanishing slope in figure \ref{actionplot} for arbitrarily small deformations. By expanding the solution for the embedding function $\theta(z)$, the deformation $A$ can be interpreted as the expectation value of a dimension one operator in the UV quantum mechanics of the boundary impurity \cite{Kumar:2016jxy}.

\begin{figure}[h]
\begin{center}
\includegraphics[width=1.9in]{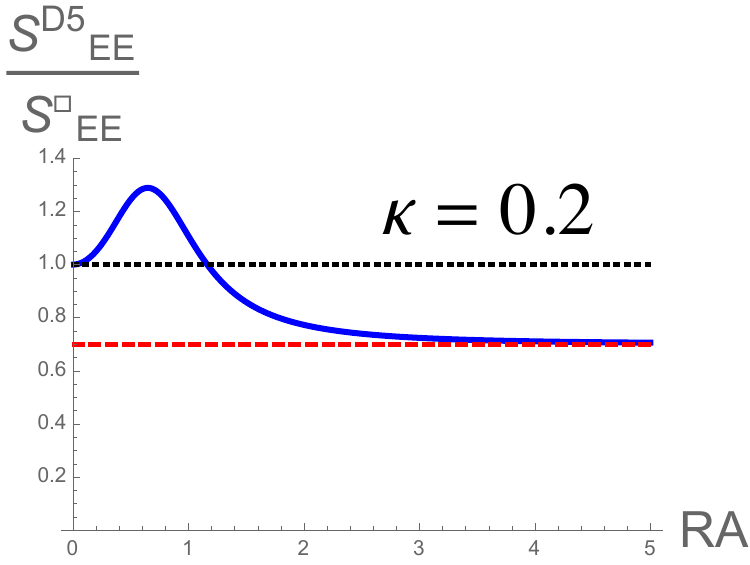}
\includegraphics[width=1.9in]{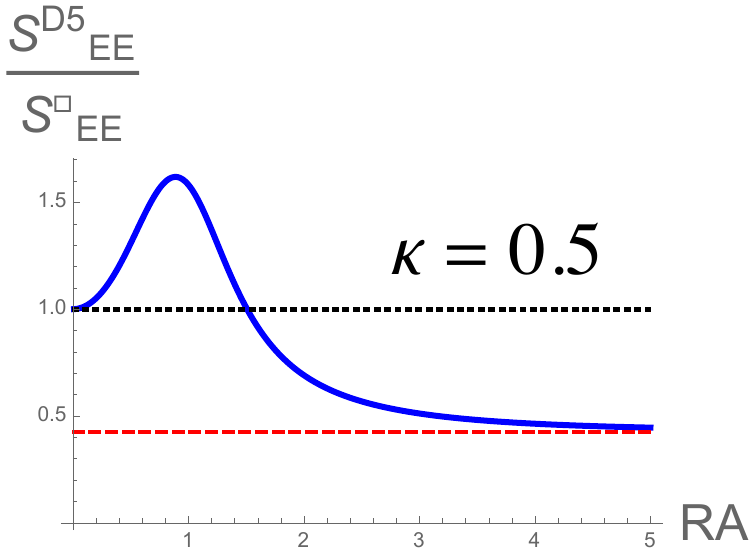}
\includegraphics[width=1.9in]{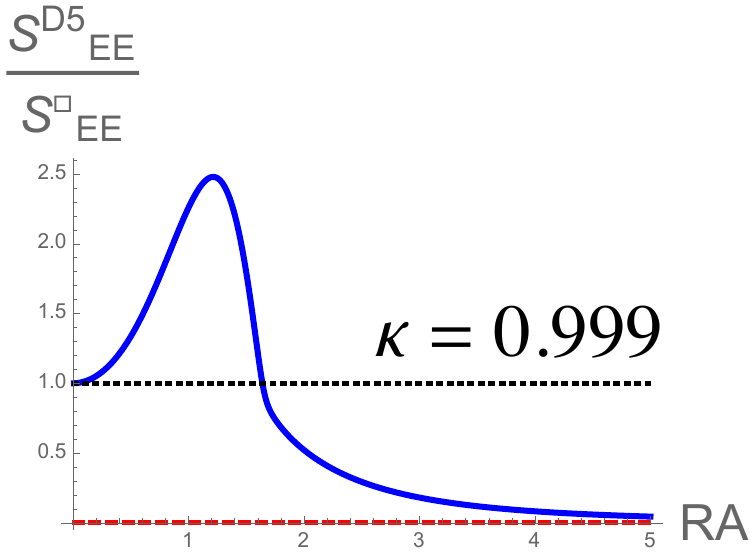}
\caption{\small{The ratio of the EE contribution due to the D5-brane impurity (thick blue) and that of $k$ fundamental quarks (dotted black line), as a function of the radius $R$ of the spherical entangling region. The deformation $A$ drives the flow and the dimensionless tunable parameter is $RA$. The red dashed line is the ratio $S_{{\cal A}_k}/(kS_\Box)$.}}
\label{figd5flow}
\end{center}
\end{figure}

The EE contribution from the impurity is obtained by varying the ``off-shell" action \eqref{offshellac} with respect to $\beta$ and evaluating the first variation on the BPS solution,
\bea
&&S_{\rm D5}(RA)\,=\,\lim_{\beta\to 2\pi}\beta\,\partial_\beta {\rm I}_{\rm D5}(\beta)\\\nonumber
&&=\,{\rm T}_{\rm D5}\,{\rm Vol}({\rm S}^4)\left[\frac{\pi}{3}\left.\frac{\partial_\theta\tilde D}{\tilde D}\right|_{\rho=1}\,+\,\frac{1}{3}\int_{-\frac{\pi}{2}}^{\frac{\pi}{2}}d\tau\int_{1}^\infty d\rho \frac{(\partial_\theta \tilde D)^2}{\tilde D}\,\frac{1\,-\,2 z^2\,\sin^2\tau/R^2}{\rho^2(\rho^2-1)}
\right]\,.
\eea
We have made use of the BPS formula \eqref{bps} and that $\beta\partial_\beta \rho_+\,=\,-\frac{1}{3}$ when $\beta=2\pi$. 
Recasting the result in terms of the integral over the domain ${\cal D}: x_0^2\,+\,z^2 \leq R^2$ in Poincar\'e patch, we find:
\bea
&&S_{\rm D5}(RA)\,=\,\lim_{\beta\to 2\pi}\beta\partial_\beta {\rm I}_{\rm D5}(\beta)\,\\\nonumber\\\nonumber
&&= {\rm T}_{\rm D5}\frac{8\pi^2}{3}\left[\frac{\pi}{3}\left.\frac{\sin^8\theta\,+\,D^2}{\sin^5\theta\,+\,D\,\cos\theta}\right|_{\rho\,=\,1}\,-\,\frac{1}{3}\int dx_0\int dz\,\theta'(z)\,\sin\theta\,\left(\sin^3\theta\cos\theta\,-\,D\right)\times\right.\\\nonumber\\\nonumber
&&\qquad\qquad\qquad\qquad\qquad\left.\times\frac{16 R^4\, z^3 \,\left(x_0^4\,+\,x_0^2(2 R^2-6z^2)+(z^2-R^2)^2\right)}{(z^2+x_0^2+R^2)^2\,((x_0^2+z^2)^2+2R^2(x_0^2-z^2)+R^4)^2}\right]\,.
\eea
As in the case of the free energy above, the integration over the domain ${\cal D}$  must be performed numerically.  The integral  over the $x_0$ coordinate can once again be obtained analytically, and the final integration is achieved numerically after exchanging $z$ for $\theta$. The result for the entanglement entropy excess is a function of the dimensionless combination $(RA)$, as plotted in fig.\eqref{figd5flow}.
For every value of $\kappa=k/N$, we see that the entanglement entropy contribution interpolates between that of $k$ coincident fundamental quarks and a source in the antisymmetric representation ${\cal A}_k$:
\be
\left.\frac{S_{\rm D5}}
{k\,S_{\,\Box}}\right|_{AR \to 0}\,=\,1\,,\qquad\qquad \left.\frac{S_{\,\rm D5}}
{k\,S_{\,\Box}} \right|_{AR \to\infty}\,=\,\frac{2}{3\pi\kappa}\,\sin^3\theta_{\kappa}\,.
\ee
The main notable feature of the results is that the variation of the EE with size of the entangling region (or equivalently the deformation $A$) is non-monotonic, exhibiting a maximum at a special value of $AR$ of order unity, and decreasing monotonically subsequently.  

\subsection{Comparison with $\langle {\cal O}_{F^2}\rangle$}
\begin{figure}[h]
\begin{center}
\includegraphics[width=2in]{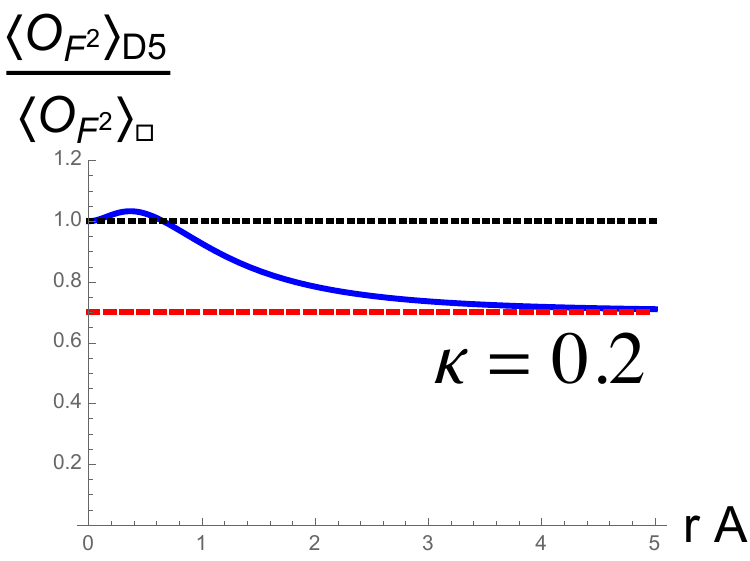}\includegraphics[width=2in]{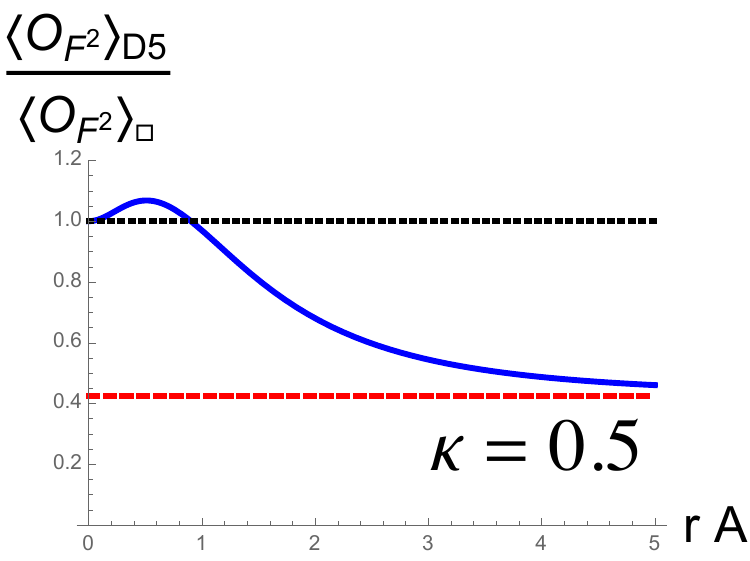}
\includegraphics[width=1.9in]{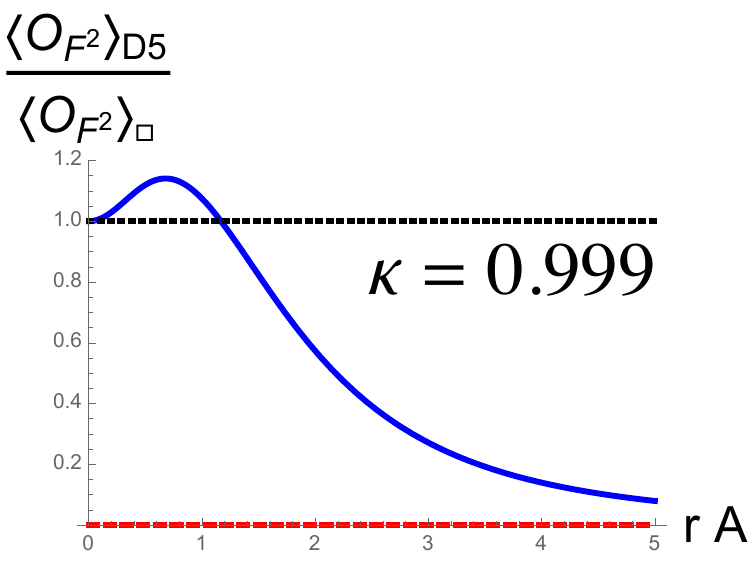}
\caption{\small{The VEV of the dimension four operator ${\cal O}_{F^2}$ dual to the dilaton, in the presence of the non-conformal impurity, divided by the corresponding VEV due to $k$ coincident quarks (dotted black), plotted (solid blue) as a function of $(r A)$. The red dashed line denotes the VEV of the same operator in the presence of the source in representation ${\cal A}_k$.}}\label{vev}
\end{center}
\end{figure}

The D5-brane is a source of various supergravity fields in ${\rm AdS}_5\times {\rm S}^5$ and the falloffs of these fields yield the VEVs of  corresponding operators in the boundary gauge theory. In particular, the dilaton falloff was used in \cite{Kumar:2016jxy} to infer the VEV of the dimension four operator ${{\cal O}_F^2}\,=\,{\rm Tr}F^2\,+\ldots$, equal to the Lagrangian density of the ${\cal N}=4$ theory, in the presence of the non-conformal D5-brane 	impurity. 
Since ${\cal O}_{F^2}$ is a dimension four operator, for conformal impurities the VEV of this operator scales as $1/r^4$ where $r$ is the spatial distance from the heavy quark on the boundary:
\bea
\langle{\cal O}_{F^2}\rangle &&=\,\frac{\sqrt 2}{24\pi^2}\,\left(\frac{3\pi \kappa}{2}\right)\,\frac{\sqrt\lambda}{r^4}\,,\qquad\qquad
r A \ll 1\,,\\\nonumber\\\nonumber
&&=\,\frac{\sqrt 2}{24\pi^2}\sin^3\theta_\kappa\,\frac{\sqrt\lambda}{r^4}\,,
\qquad\qquad\quad r A \gg 1\,.
\eea
In fig. \eqref{vev}, we plot the dimensionless ratio $\langle {\cal O}_{F^2}\rangle_{\rm D5}/\langle {\cal O}_{F^2}\rangle_{\Box}$ as a function of the dimensionless distance from the impurity $(r A)$. 
The qualitative features of the plots are similar to those of the entanglement entropy contribution from the defect.  The sources in the fundamental representation are screened into the antisymmetric representation, but the effect is non-monotonic as a function of the distance from the source.
 
\section{D3-brane impurities}
The D3-brane embedding with worldvolume ${\rm AdS}_2\times {\rm S}^2 \subset {\rm AdS}_5$ found by Drukker and Fiol \cite{drukkerfiol} computes BPS Wilson lines in the rank $k$ symmetric tensor representation ${\cal S}_k$ \cite{passerini,  paper1}.  In \cite{Kumar:2016jxy}, a D3-brane (BPS) embedding was analyzed which interpolates between the representation ${\cal S}_k$ in the UV and $k$ coincident strings in the IR. We will first review the properties of this zero temperature solution in Poincar\'e patch and subsequently analyze its geometric entropy. 

\subsection{Poincar\'e patch D3-brane embedding}
The D3-brane wraps an AdS$_2\times {\rm S}^2$ subset of AdS$_5$ and is supported by $k$ units of flux. Since the internal five-sphere plays no role we will suppress it in the discussion below. The D3-brane impurity preserves the same symmetries as a point at the spatial origin of the boundary CFT on ${\mathbb R}^{1,3}$. In particular, choosing  the worldvolume coordinates to be  $(x_0, \sigma, \Omega_2)$ the induced metric for the relevant embedding takes the form (in Euclidean signature),
\be
*ds^2\Big|_{\rm D3}\,=\,\frac{1}{z^2}\left[\,dx_0^2\,+\,{d\sigma^2}\left[\left(\frac{\partial z}{\partial \sigma}\right)^2\,+\,\left(\frac{\partial r}{\partial \sigma}\right)^2\right]\,+\,r(\sigma)^2\,d\Omega_2^2 \right]\,.
\ee
Eventually we will set $\sigma = z$ after discussing the counterterms and UV regularization.  The background five-form RR flux, and its associated four-form potential play a crucial role in stabilizing the D3-brane configuration. In particular, the pullback of the four-form potential onto the D3-brane worldvolume is
\be
*C_4\,=\,-\frac{i}{g_s}\,\frac{r^2}{z^4}\,\partial_\sigma r\, dx_0\wedge d\sigma\wedge \omega_2\,,\label{c4straight}
\ee 
where $\omega_2$ is the volume-form on the unit two-sphere.  We also recall  that $C_4$ is only defined up to a gauge choice. The choice of gauge will be important when we proceed to the calculation of the entanglement entropy contribution from the defect.  The expanded D3-brane configiuration also has 
a worldvolume electric field $G\,=\,2\pi i \alpha'\,F_{0\sigma}$ and the a tension $T_{\rm D3} = \frac{N}{2\pi^2}$. Putting all ingredients together, we find, 
\bea
{\rm I}_{\rm D3}\,=&&T_{\rm D3}\int dx_0\,d^3\sigma\,
\sqrt{{\rm det}\left(*g\,+\,2\pi\alpha' F\right)}\,-\,ig_s\int *C_4\,+\,{\rm I}_{\rm c. t.}
\label{d3poincare}\\\nonumber
\\\nonumber
=&&\frac{2N}{\pi}\int dx_0\,d\sigma\,\frac{r^2}{z^4}\left(\sqrt{(\partial_\sigma r)^2\,+\, (\partial_\sigma z)^2\,-\,G^2\,z^4}\,-\,\partial_\sigma r\right)\,+\,{\rm I}_{\rm c.t.}\,.
\eea
As in the D5-brane case, the counterterms consist of a piece that fixes the number of units of the F-string charge to be $k$ and another that exchanges Dirichlet boundary conditions for Neumman ones for the active worldvolume field(s) in the embedding,
\bea
&& I_{\rm c.t.}\,=\, {\rm I}_{\rm U(1)}\,+\,{\rm I}_{\rm UV}\,,\qquad\qquad
{\rm I}_{\rm U(1)}\,=\,\frac{k\sqrt\lambda}{2\pi} \int dx_0\,d\sigma\,G\,,\\\nonumber
\\\nonumber
&& {\rm I}_{\rm UV}\,=\,-\int dx_0\left[r\,\frac{\delta I}{\delta(\partial_\sigma r)}\,+\,z\,\frac{\delta S}{\delta (\partial_\sigma z)}\,\right]_{\rm UV}\,.
\eea
The equations of motion for $G$ and for $r(z)$ (after picking the gauge $\sigma =z$) \cite{Kumar:2016jxy} are solved by BPS configurations satisfying the first order equations
\bea
\frac{\partial r}{\partial z}\,=\, G\,\frac{r^2}{\tilde \kappa}\,,\qquad G\,=\,\frac{1}{z^2}\,,\qquad \tilde \kappa\,\equiv\,\frac{k\sqrt\lambda}{4N}\,.
\eea
The most general solution to the first order equation is,
\be
r\,=\,\frac{\tilde \kappa\, z}{1\,+\,a \tilde\kappa \, z}\,,\label{d3bps}
\ee
where $a$ is a constant of integration.  The solution with $a=0$ is the expanded D3-brane solution of \cite{drukkerfiol} which describes the BPS Wilson line in the symmetric tensor representation ${\cal S}_k$. The  configuration with $a>0$ on the other hand is a deformation of the worldline theory on the corresponding heavy quark  which drives a flow from the representation ${\cal S}_k$ to $k$ coincident quarks. Finally, the  solution with $a<0$ can naturally be interpreted as a 
heavy quark source on the Coulomb branch \cite{Schwarz:2014rxa} of the ${\cal N}=4$ theory with $SU(N)$ broken to $U(1)\times SU(N-1)$ by displacing one of  $N$ coincident D3-branes  and placing it at $z= -1/a\tilde\kappa$. The brane has a soliton/lump due to a W-boson state corresponding to a string or a bundle of strings stretching from the displaced D3-brane to the conformal boundary at infinity. It appears in the UV as a quark source transforming in the representation ${\cal S}_k$ of $SU(N)$.

\section{D3-brane entanglement entropy}
\subsection{Choice of $C_4$}
The calculation of EE for any of the D3-brane configurations described above is complicated by  the contribution from the pullback of the RR four-form potential. 
To understand this statement we recall that for conformal impurities, the EE contribution has been argued to be determined by the value of the circular BPS  Wilson loop \cite{Lewkowycz:2013laa}. It was noted first in \cite{drukkerfiol} that the calculation of the circular Wilson loop using D3-branes requires a different choice of gauge for the RR four-form potential than the one used (cf. eq.\eqref{c4straight}) to obtain the straight Wilson line. The circular Wilson loop  is related to the straight one by a conformal transformation. Extension of this into the bulk  AdS yields the coordinate transformation \eqref{specialconformal}   from (Euclidean) Poincar\'e patch AdS to the Rindler-AdS  metric \eqref{rindlerads}. Crucially, if one simply uses the transformed version of the RR four-form potential \eqref{c4straight} one does not obtain the correct result for the circular Wilson loop. Instead one must use the natural form for the four-form potential in (Euclidean) Rindler-AdS:
\bea
&&ds^2_{\rm E}\,=\,\frac{1}{\tilde z^2}\left(d\tilde z^2\,+\,dr_1^2\,+\,r_1^2d\tau^2\,+\,dr_2^2\,+\,r_2^2 d\phi^2\right)\,,\label{c4adshyp}\\\nonumber\\\nonumber
&& C_4\big|_{{\rm Rindler-\rm AdS}}\,=\,-\frac{i}{g_s}\frac{r_1 r_2}{\tilde z^4}\,dr_1\wedge d\tau\wedge dr_2\wedge
d\phi\,,
\eea
and transform these to hyperbolic AdS using eq.\eqref{rindlertohyp}.  This procedure was shown to yield the result for the circular Wilson loop \cite{drukkerfiol} in the representation ${\cal 
S}_k$.
It was also noted in \cite{drukkerfiol}, that  $C_4$ in eq. \eqref{c4adshyp} is gauge equivalent to  the corresponding expression \eqref{c4straight} in Poincar\'e patch  so that $\left.C_4\right|_{\rm Rindler-AdS}\,=\,\left.C_4\right|_{\rm Poincare'}\,+\, d\Lambda_{3}$. Upon transforming to hyperbolic AdS coordinates, 
the above four-form potential reads,
\bea
&&i\,g_s\,C_4\,  =\label{D3C4}\\\nonumber
&& \rho^2(\rho^2-1)\sinh^2u\sin\vartheta\, du\wedge d\tau\wedge d\vartheta\wedge d\varphi \,+\,\fr{\rho\sinh u\sin^2\vartheta}{\cosh u -\cos\vartheta\sinh u} d\rho\wedge d\tau\wedge du \wedge d\varphi \\\nonumber\\\nonumber
&&-\,\frac{\rho\sinh^2u\sin\vartheta(\sinh u \,-\,\cos \vartheta\cosh u)}{\cosh u-\cos\vartheta\sinh u}d\rho\wedge d\tau \wedge d\vartheta \wedge d\varphi\,,
\eea
where  $(\vartheta, \varphi)$ are standard angular coordinates on the spatial two-sphere. The term proportional to $\rho^4$ is the natural four-form potential on the hyperbolically sliced AdS background whose exterior derivative yields the volume form (the five-form flux $F_5$) on AdS$_5$.  The remaining terms can be shown explicitly to combine and reduce to a pure gauge transformation. Importantly, these must be retained in order to obtain the correct result for the D3-brane circular Wilson loop.

The BPS embedding \eqref{d3bps} for the D3-brane in hyperbolic AdS coordinates is 
\be
\rho\sinh u\,=\,\tilde\kappa\,\frac{\rho\cosh u\,+\,\sqrt{\rho^2-1}\,\cos\tau}{\tilde \kappa\,(aR)\, + \,\rho\cosh u\,+\,\sqrt{\rho^2-1}\,\cos\tau}\,,\label{hypadsbps}
\ee
and can be viewed as specifying $u$ as a function $u(\tau,\rho)$ of the radial AdS coordinate $\rho$ and Euclidean time $\tau$. The undeformed, conformal solution is obtained when $aR \to 0$, yielding $\rho\sinh u\,=\,\tilde\kappa$. The pullback of $C_4$ onto the worldvolume of this embedding contains only two of the three terms in \eqref{D3C4}:
\bea
&&*C_4\,=\\\nonumber
&&\,\left[\rho^2(\rho^2-1)\,\partial_\rho u\, -\,\frac{\rho(\sinh u \,-\,\cos \vartheta\cosh u)}{\cosh u-\cos\vartheta\sinh u}\right]\sinh^2 u\sin\vartheta\, d\rho\wedge d\tau \wedge d\vartheta \wedge d\varphi\,.
\eea
Then the Wess-Zumino term of the D3-brane action, upon integration over the spatial $S^2$ yields:
\bea
{\rm I}_{\rm WZ}\,=&&-ig_s {\rm T}_{\rm D3}\int *{C_4}\label{type1WZ}\\\nonumber\\\nonumber
=&&\frac{N}{2\pi}\int_{-\frac{\pi}{2}}^{\frac{\pi}{2}} d\tau\int_{1}^\infty d\rho\, \left[\rho^2(\rho^2-1)\,\sinh^2u\,\partial_\rho u\,+\,\rho(u\,-\,\sinh u\cosh u)\right]\,.
\eea
Note that we have only made use of the fact that $u=u(\rho,\tau)$,  without using the explicit form of the BPS solution. 

The next question we must ask is whether $C_4$ needs to be modified when the temperature $\beta$ of the hyperbolic black hole is different from $2\pi$, an issue which will become relevant when we implement the replica trick. Any modification in $C_4$ can only be pure gauge, and such choice of gauge will require independent justification.
The simplest assumption is that $C_4$ remains unchanged even with $\beta\neq 2\pi$. This is natural, but we will see that this approach leads to a result for the entanglement entropy in disagreement with \cite{Lewkowycz:2013laa} which relates the EE contribution to the circular Wilson loop via eq.\eqref{EEcirc} for conformal impurities (the $a=0$ embedding). %However we will also find that this proposal for EE actually tracks the strength of the heavy quark source measured by the expectation value of the operator ${\cal O}_F^2$ as in \cite{Kumar:2016jxy}, and decreases monotonically from the UV to the IR in the situations with $a\neq 0$

We  propose a simple modification to $C_4$  when $\beta\neq 2\pi$. This modification needs to be pure gauge {\em and} temperature dependent in just the right way so as to reproduce the entanglement entropy result for the symmetric representation as predicted by \cite{Lewkowycz:2013laa}. Importantly, we would like it to only have support at the locus of points where the $\tau$-circle shrinks, namely at the hyperbolic horizon $\rho=\rho_+$ \eqref{Thawk}, when $\beta\neq 2\pi$. Based on these criteria we find that the shift,
\be
C_4\to C_4\,-\,{\cal F}(\rho_+)\, \sinh^2 u\, \sin\vartheta \, du\wedge d\tau\wedge d\vartheta\wedge d\phi\,,
\ee
with
\be
{\cal F}(1)\,=\,0\,,\qquad\qquad \partial_\beta{\cal F}(\rho_+)\big|_{\rho_+=1}\,=\,-\frac{2}{3}\,,
\ee
satisfies all requirements. The precise dependence on $\rho_+$ is not important. The function must vanish when $\rho_+=1$ (or $\beta=2\pi$) and its first derivative is constrained by matching to the entanglement entropy for the conformal impurity. For concreteness, we take 
\be
{\cal F}(\rho_+)\,=\,\rho_+^2(\rho_+^2-1)\,,
\ee
because it has the effect of modifying the relevant component of $C_4$ in a natural way,
\be
\rho^2(\rho^2-1)\sinh^2 u \,du\wedge d\tau\wedge d\omega_2 \,\to\,
\rho^2\,f_{n}(\rho)\,\sinh^2 u \,du\wedge d\tau\wedge d\omega_2\,.\label{c4shift}
\ee
%\be
%ds_{\rm D3}^2\,=\,\left(\frac{1}{v^2}\left(\frac{\pa v}{\pa r}\right)^2+v^2\right)dr^2+v^2(dt^2+r^2d\Omega_2^2) \pm 2drdtG
%\ee
%\be
%r^2v^4\Omega_2\,=\, \int  d^{\, 4}x\,C_{(4)},\quad -\k\,=\, \frac{\d I}{\d G},\quad G\,=\,\frac{\k}{r^2}
%\ee
%\be
%I\,=\,T\int_{-\infty}^{\infty}dt\int_0^{\infty}dr\,%\left[ r^2v^2\sqrt{(\pa_rv)^2+v^4-G^2}-r^2v^4+G\k\right]-I_{c.t.}
%\ee
%\be
%v\,=\,\frac{\k}{r},\qquad I_{c.t.}\,=\,\lim_{r\to 0}\,\frac{\k}{r},\qquad I\,=\,0 
%\ee
\subsection{D3-brane action in hyperbolic AdS}
The D3-brane solutions with $a\neq 0$ are non-static embeddings in hyperbolic AdS, given by eq.\eqref{hypadsbps}, so that $u=u(\tau,\rho)$. Formally, the action for these embeddings evaluated in the hyperbolic AdS black hole geometry with generic $\beta$ is, 
\bea
{\rm I}_{\rm D3}(\beta)\,=&&\frac{2N}{\pi}\int_{-\frac{\pi}{2}}^{\frac{\pi}{2}}d\tau\int_{\rho_+}^\infty d\rho\left[\rho^2\sinh^2u\sqrt{1\,+\, \rho^2f_n(\rho) (\partial_\rho u)^2\,+\,\rho^2
\frac{(\partial_\tau u)^2}{f_n(\rho)}\,-\,G^2 }\right.\nonumber\\\nonumber\\
&&\left.+\,\rho^2\,f_{\rm WZ}(\rho)\,\sinh^2u\,\partial_\rho u\,+\,\rho(u\,-\,\sinh u\cosh u)\,+\, G\,\tilde \kappa\right]\,+\,{\rm I}_{\rm UV}\,.\label{d3actionhyp}
\eea
We have introduced the function $f_{\rm WZ}(\rho)$ which includes a slight generalization of the pure gauge shift \eqref{c4shift}:
\be
f_{\rm WZ}(\rho)\,=\, \rho^2-1\,-\,\gamma\,\frac{\rho_+^2(\rho_+^2-1)}{\rho^2}\,,
\ee
 where $\gamma$ can be treated as a free parameter, so that we may see how different choices of $\gamma$ affect the final results.
%\be
%\k\,=\,\r\sinh u,\quad G\,=\,\frac{\r}{\sqrt{\k^2+\r^2}}
%\ee 
%\be
%I\,=\, T\int_{0,1}^{2\pi,\infty}d\t d\r\,\r^2\sinh^2u\sqrt{1+\r^2(\r^2-1)(\pa_\r u)^2-G^2}+I_{\rm W.Z.}+G\k+I_{\rm c.t.}
%\ee
Note that this shift in $C_4$ is pure gauge for any value of $\gamma$ and vanishes at $\beta=2\pi$.
The equation of motion for the electric field $G$ is algebraic. Solving for it and substituting the result into the D3-brane action, we obtain,
\bea
{\rm I}_{\rm D3}(\beta)\,=&&\frac{2N}{\pi}\int_{-\frac{\pi}{2}}^{\frac{\pi}{2}}d\tau\int_{\rho_+}^\infty d\rho\left[\sqrt{\left(\kappa^2\,+\,\rho^4\sinh^4u\right)\left(1\,+\, \rho^2f_n(\rho) (\partial_\rho u)^2\,+\,\rho^2
\frac{(\partial_\tau u)^2}{f_n(\rho)} \right)}\right.\nonumber\\\nonumber\\
&&\left.+\rho^2\,f_{\rm WZ}(\rho)\,\sinh^2u\,\partial_\rho u\,+\,\rho(u\,-\,\sinh u\cosh u)\right]\,+\,{\rm I}_{\rm UV}\,.\label{d3aceval}
\eea 
\subsection{Conformal D3-embedding: symmetric representation }
We will first  rederive  results 
for the action and the entanglement entropy of the conformal, or $a=0$ embedding. 
The straight Wilson line in the symmetric representation is given by the Poincar\'e patch embedding $r\,=\,\tilde\kappa\, z$, which after transforming to hyperbolic AdS coordinates, 
becomes,
\be
\tilde\kappa\,=\,\rho\sinh u\,.
\ee
It is fairly easy to check that this embedding solves the equations of motion following from the action  \eqref{d3actionhyp} with $\beta=2\pi$, treating $G$ and $u$ as independent degrees of freedom. Plugging this solution into eq. \eqref{d3aceval}, we find that the  Born-Infeld and 
Wess-Zumino terms {\em almost} cancel out each other at the level of the integrands, leaving behind only the contribution linear in $u$, so that
\be
{\rm I}_{\rm D3}(2\pi)\big |_{a=0}\,=\, 2N\int_1^{\rho_\infty} d\rho\,\rho\,\sinh^{-1}\left(\frac{\kappa}{\rho}\right)\,+\,{\rm I}_{\rm UV}\,.
\ee
Following the procedure described earlier, the UV counterterm is,
\be
{\rm I}_{\rm UV}\,=\,\left.\rho\frac{\d I}{\d (\partial_u\rho)}\right|_{\rho=\rho_\infty\gg 1}\,=\,-\kappa\,\rho_\infty\,.
\ee
Performing the integrals and subtracting off the divergent piece against the UV counterterm, we obtain the well known result of \cite{drukkerfiol} for the circular Wilson loop,
\be
{\rm I}_{\rm D3}(2\pi)\big|_{a=0}\,=\, - N\left(\tilde \kappa\sqrt{1\,+\,\tilde \kappa^2}\,+\,\sinh^{-1}\tilde \kappa\right)\,.
\ee
%\begin{center}
%\includegraphics[height=6cm]{D3loop}
%\end{center}
%\textbf{Figure 3}: Displays the free energy, as a function of the rank of the repSresentation, for the totally symmetric circular Wilson loop.
Now, we turn to the entanglement entropy contribution due to the impurity in the symmetric representation ${\cal S}_k$.  Differentiating the off-shell action ${\rm I}_{\rm D3}(\beta)$ with respect to $\beta$, we find for the conformal ($a=0$) embedding:
\bea
S_{{\cal S}_k}\,=&&\lim_{\beta\to2\pi} \beta\,\partial_\beta {\rm I}_{\rm D3}(\beta)\\\nonumber\\\nonumber
=&&N\left(\tfrac{2\gamma+1}{3}\sinh^{-1}\tilde\kappa\,-\,\tfrac{2\gamma-1}{3}\tilde \kappa\sqrt{1+\tilde \kappa^2}\right)\,.
\eea
Setting $\gamma=1$ we obtain precisely the expression for the EE associated to symmetric tensor representation \eqref{EElist}.  This fixes the choice of gauge to be as given in eq.\eqref{c4shift}. Interestingly, the limit of small $\tilde\kappa $ is actually independent of $\gamma$ and yields the EE entropy associated to $k$ fundamental strings:
\be
S_{{\cal S}_k} \,\simeq\,k\frac{\sqrt\lambda}{6}\,=\,kS_\Box\,,\qquad \tilde\kappa \ll 1\,.
\ee
Thus, the value of $\gamma$ could not have been fixed by matching to the result for $k$ fundamental quarks in the limit of small $\tilde\kappa$.

\subsection{Action on $S^1_\beta\times H^3$ with deformation $a > 0$}
Having identified the appropriate gauge in which the RR four-form yields the correct EE for the symmetric representation, we turn to the calculation for the non-conformal  solution with $a\neq 0$.  We first evaluate the action for the D3-brane flow solution when mapped to the hyperbolic AdS geometry. This yields  the free energy of the impurity on $S^1_\beta\times H^3$ at a temperature $\beta^{-1} = \frac{1}{2\pi}$.  If evaluated directly on the Poincar\'e patch embedding \eqref{d3poincare} restricted 
to the domain ${\cal D}: z^2+r^2+x_0^2 \leq R^2$, the result does not match the circular Wilson loop\footnote{In this case the DBI and Wess-Zumino terms cancel at the level of the Langrangian densities, leaving behind only the counterterms ${\rm I}_{\rm c.t.}\,=\,\frac{2N}{\pi}\int \int_{\cal D}dx_0 dz\,\tilde\kappa G\,+\,{\rm I}_{\rm UV}$.  For the conformal embedding $r=z\tilde\kappa$, the integration is simple, and the result does not match the circular Wilson loop.}. Therefore, it is necessary to first formulate the calculation in the hyperbolic AdS embedding with the correct gauge choice for $C_4$. 

In appendix \ref{appa},  we show the steps involved in computing the D3-brane action by first writing it in the Euclidean hyperbolic AdS background, and subsequently translating it to an integral over the worldvolume restricted to the domain ${\cal D}$ in Poincar\'e patch:
\be
{\cal D}:  \quad x_0^2 \,+\,z^2\,+\,r(z)^2\,\leq R^2\,,\qquad
r(z)\,=\,\frac{\tilde\kappa z}{1+ a\tilde\kappa z}\,.
\ee
We find that the action for the non-conformal impurity placed in hyperbolic space $H^3$ at $\beta=2\pi$ is given by the expression,
\bea
&&{\rm I}_{\rm D3}(2\pi)\,=\,\frac{2N}{\pi}\int\int_{\cal D}dx_0\,dz\,\left[\frac{\tilde \kappa}{z^2}\,-\,\frac{a \tilde \kappa^4}{z(1+az\tilde\kappa)^4} +\,\frac{a\tilde\kappa^4}{z(1+az\tilde\kappa)^4}\times\right.\label{d3actionint}\\\nonumber\\\nonumber
&&\left.
\left\{\frac{\left((x_0^2+R^2)^2-z^4+2 r(z)^2(x_0^2-R^2)+r(z)^4\right)}{r(z)^4+2r(z)^2(x_0^2+z^2-R^2)+(x_0^2+z^2)^2+2R^2(x_0^2-z^2)+R^4}\right\}\right]\,+\,\\\nonumber\\\nonumber
&& -\,N(u_+ \,-\,\sinh u_+ \cosh u_+)\,-\,4N\tilde\kappa \frac{R}{\epsilon}\,.
\eea
Here $u_+$ is the value  \eqref{uplus} of $u(\rho,\tau)$ at the hyperbolic AdS black hole horizon $\rho=1$ where the Euclidean temporal circle parametrized by $\tau$, shrinks smoothly. The expression satisfies some immediate checks. For vanishing $a$, only the first term of the integrand above survives and we obtain the result for the Wilson loop in the symmetric representation  ${\cal S}_k$ upon integrating over the domain ${\cal D}$ and including the horizon contribution depending on $u_+$. In the opposite limit of large $a$, the first term in the integrand dominates once again and reduces to the action for $k$ strings. Another interesting feature of the integrand  is that it contains  $a$-dependent terms which individually produce logarithmic divergences at small $z$, but these cancel precisely against each other ensuring that the the UV divergence structure is unaltered.
\begin{figure}[h]
\begin{center}
\includegraphics[width=3in]{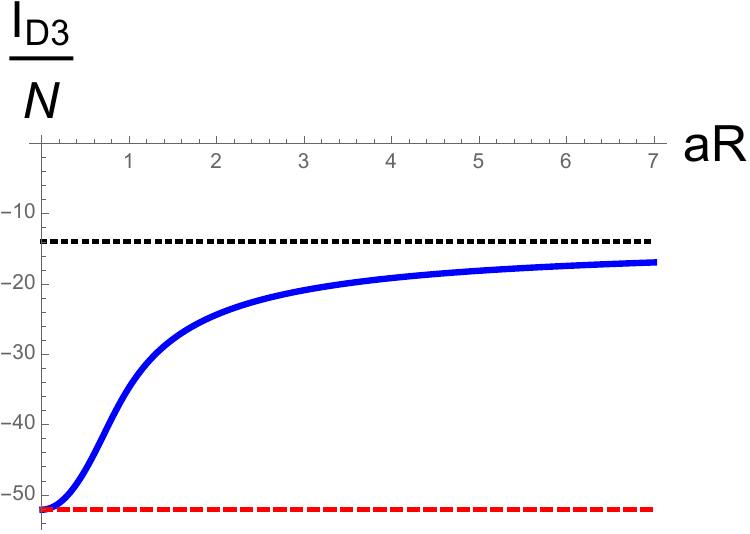}
\end{center}
\caption{\small{The free energy $-\ln Z_{\cal H}\,=\,{\rm I}_{\rm D3}(2\pi)$  of the D3-brane impurity in hyperbolic space with $\beta=2\pi$ and $\tilde\kappa =7.0$ increases towards the IR as the deformation $(a R)$ is increased. In the UV it matches the symmetric Wilson loop (dashed red line) and asymptotes in the IR towards $k$ times the circular Wilson loop in the fundamental representation.}}
\label{d3flowaction}
\end{figure}
Figure \ref{d3flowaction} shows that the free energy ${\rm I}_{\rm D3}(2\pi) < 0$ increases monotonically as a function of $(aR)$ and smoothly connects the symmetric representation (UV) to $k$ coincident quarks in the IR. Analogously to the D5-brane case, this points to an interpretation in terms of a relative entropy, since the deformation $a$ corresponds to an expectation value of a dimension one operator in the UV \cite{Kumar:2016jxy}, and can be viewed as labelling a state different from the thermal one. 
\subsection{Action for D3-brane embedding with $a<0$}
\begin{figure}[h]
\begin{center}
\includegraphics[width=2.7in]{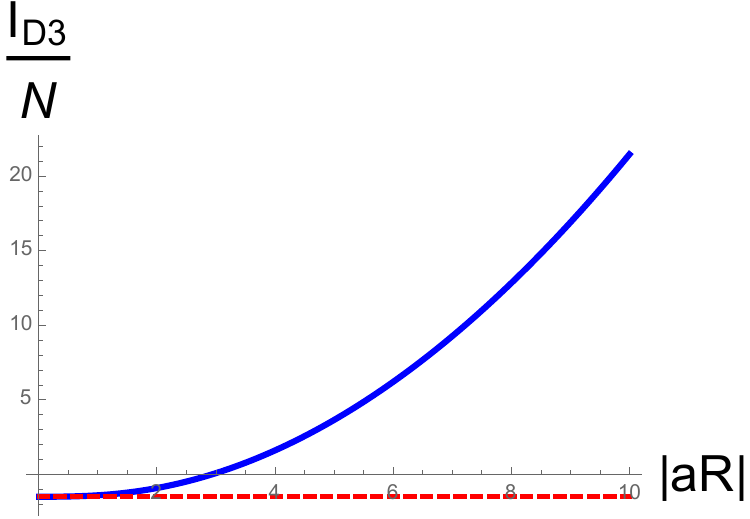}\hspace{0.2in}
\includegraphics[width=2.7in]{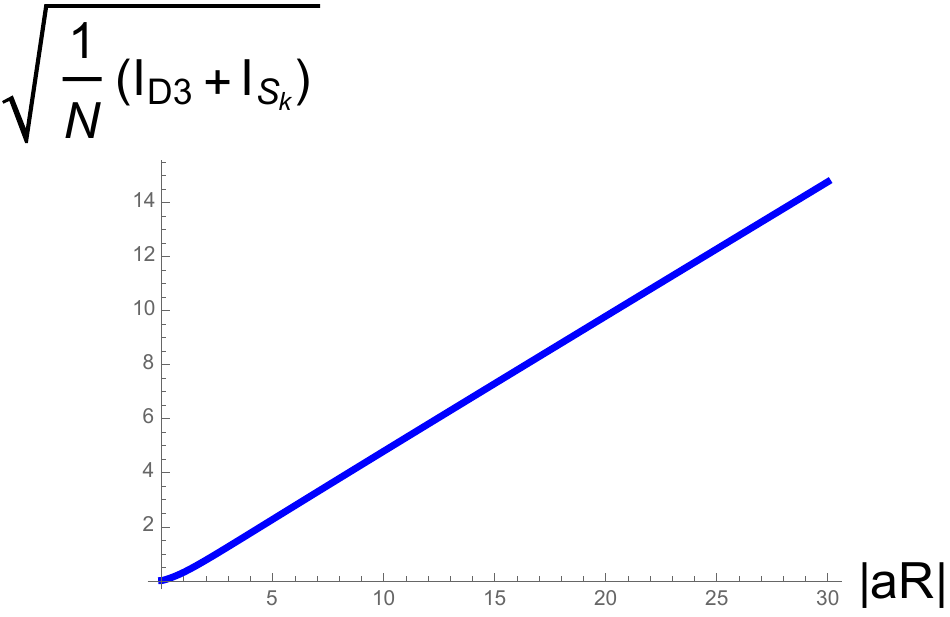}
\end{center}
\caption{\small{The free energy $-\ln Z_{\cal H}\,=\,{\rm I}_{\rm D3}(2\pi)$  of the D3-brane solution with $a<0$ in hyperbolic space with $\beta=2\pi$ and $\tilde\kappa =0.7$. 
For small deformations (UV), it matches the symmetric Wilson loop, but as $|aR|$ is increased, the free energy  increases, becomes positive and scales as $(aR)^2$.}}
\label{anegative}
\end{figure}

The D3-brane embedding with $a<0$ can be interpreted as a heavy quark   on the Coulomb branch of the ${\cal N}=4$ theory where $SU(N)$ is broken to $U(1)\times SU(N-1)$. In particular, it can be viewed as the symmetric representation Wilson line evaluated on the Coulomb branch.  Since the proper size of the $S^2$ wrapped by the D3-brane diverges at $z\,=\,|a\tilde\kappa|^{-1}$,
\be
\lim_{z\to\,|a\tilde\kappa|^{-1}}\,\frac{r}{z}\,\to\,\infty\,,
\ee
it represents a flat D3-brane at $z= |a\tilde\kappa|^{-1}$ with a spike 
stretching to the AdS-boundary. As in the case of both the D5- and D3-brane embeddings with $a>0$, the action for this configuration in hyperbolic space is a monotonically increasing function of $|a R|$ (figure \ref{anegative}).
However, unlike the previous examples, for large enough $|aR|$ the free energy becomes positive and increases without bound  as $|aR|^2$ .  

\begin{figure}[h]
\begin{center}
\includegraphics[width=1.8in]{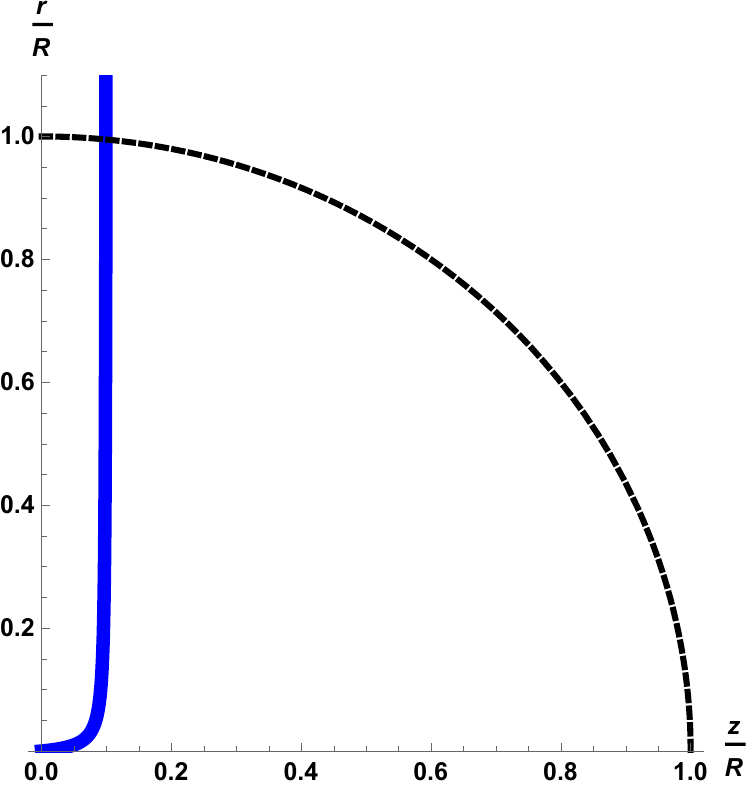}\hspace{0.1in}
\includegraphics[width=1.8in]{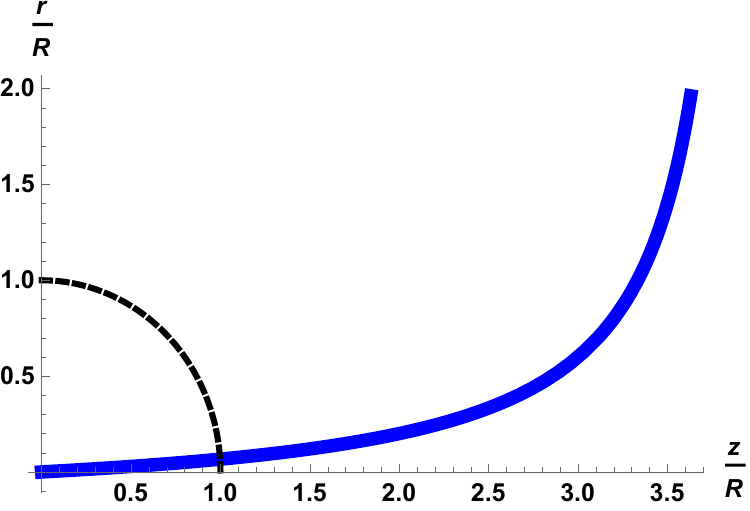}
\hspace{0.1in}
\includegraphics[width=1.8in]{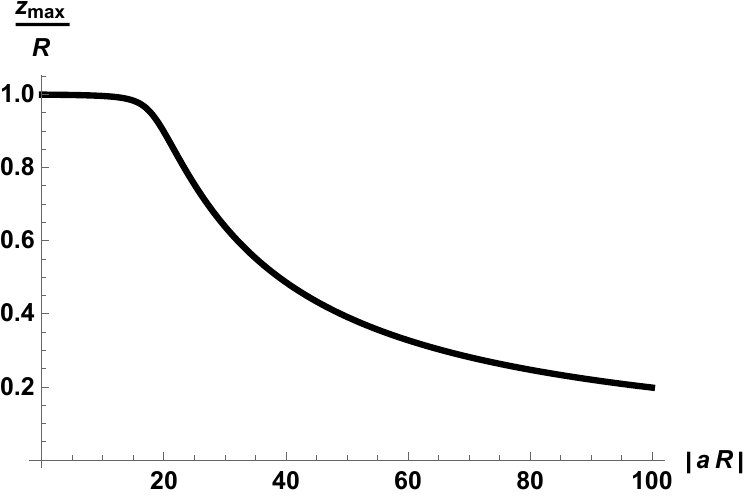}
\end{center}
\caption{\small{{\bf Left:}  Intersection of $z^2 \,+\, r^2\,=\,R^2$ with the D3-brane embedding (thick, blue curve)
for $\tilde \kappa =0.1$ and $aR=-100$. {\bf Centre:} The same
curves for $\tilde\kappa=0.05$ and $aR = -5$. {\bf Right:} Plot of the intersection point $z=z_{\rm max}$ as a function of $|aR|$ for $\tilde\kappa =0.02$. }}
\label{d3intersection}
\end{figure}

In the limit of large negative  $a$, the action   \eqref{d3actionint} can be obtained analytically and the result  
shown to agree with the numerical evaluation in figure \ref{anegative}.  The analytical approximation is based on the observation that for $|aR|\gg 1$, we are evaluating the hyperbolic space action for a Coulomb branch configuration corresponding to a D3-brane  placed at $z\,=\,|a \tilde\kappa|^{-1}$.  The situation is shown in figure \ref{d3intersection}.
For large enough $|a R|$, most of the contribution to the action is from the Coulomb branch D3-brane and the effect of the $k$ strings (in the representation ${\cal S}_k$) stretching to the conformal boundary is negligible. 
To implement this we consider the integrand in \eqref{d3actionint}, and rewrite it using  $z(r)= r/(1+ |a| r )\tilde\kappa$, so that $r$ is the independent variable. In the limit $|aR| \gg 1$, we find:
\bea
{\rm I}_{\rm D3}(2\pi)\big|_{|a R|\gg 1}\,=\,&& -\left.\,N(u_+ \,-\, \sinh u_+\,\cosh u_+)\right|_{|a R|\gg 1}\\\nonumber\\\nonumber
&&+\,\frac{N}{\pi}\int_0^ R dr \int_0^{{\sqrt{R^2-r^2}}} dx_0 \frac{2r^2 \tilde\kappa^2 |a|^2(r^2 + x_0^2 -R^2)}{r^4+2 r^2(x_0^2-R^2)+(x_0^2+R^2)^2}\,.
\eea
The first term is the horizon contribution determined by radial coordinate $u$ on the D3-brane where it intersects the hyperbolic horizon at $\rho=1$. Using the coordinate transformations \eqref{poinctohyp}, a D3-brane at $z= 
|a\tilde\kappa|^{-1}$ intersects the horizon (the boundary of the domain ${\cal D}$ in Poincar\'e patch) at,
\be
\cosh u_+\,=\, \tilde\kappa |a|R \gg 1\quad\implies\quad
u_+\,\approx\,\ln(2\tilde\kappa|a|R)\,.
\ee
We thus obtain,
\be
\left.{\rm I}_{\rm D3}(2\pi)\right|_{|aR|\gg 1}\,\simeq\,N\left(\tfrac{1}{2}(\tilde\kappa a R)^2\,-\,\ln(2\tilde\kappa|a| R)\right)\,.
\ee
This is {\em not} the entanglement entropy for the probe, but has a natural interpretation as the relative entropy of the Coulomb branch state. The  basic features are in line with the expected UV divergent contributions  to EE in four dimensional field theories \cite{Casini:2011kv} where the leading cutoff dependence is quadratic and non-universal and the subleading divergence is logarithmic with a universal coefficient.  The VEV on the Coulomb branch, given by the position of the D3-brane at $z=(|a|\tilde\kappa)^{-1}$ determines the masses of $W$-boson states and acts as a UV cutoff for the abelian factor on the Coulomb branch. The overall factor of $N$ arises due to the $N-1$ species of $W$-boson states being integrated out, viewed as open strings stretching between the single separated D3-brane and the stack of $(N-1)$ coincident branes. The coefficient (after factoring out the overall $N$) of the logarithmic contribution and its sign agrees with expected value \cite{Casini:2011kv} of $-4 a_4^*$ for a 4D CFT where $a_4^*\,=\,\frac{1}{4}$ is the A-type trace anomaly coefficient for one ${\cal N}=4$ multiplet.
Note that the gauge parameter $\gamma$ has no effect on the free energy at $\beta=2\pi$.

It is worth stressing that the logarithmic dependence originates entirely from the Wess-Zumino term of the D3-brane action from the coupling to the four-form potential, in the same gauge which yields the correct result for the circular Wilson loop. The quadratic dependence on $a$ receives contributions from both DBI and Wess-Zumino terms.

\subsection{EE for D3-brane impurity}
Now we can move on to discussing the EE contribution from the D3-brane embedding.  Differentiating the off-shell action \eqref{d3aceval} with respect to $\beta$, the entanglement entropy of the defect is,
\bea
&&S_{\rm D3}\,=\,\lim_{\beta\to 2\pi}\beta\partial_\beta {\rm I}_{\rm D3}(\beta)\label{d3eeintegral}\\\nonumber\\\nonumber
&&\frac{4N}{3\pi}\int_{-\frac{\pi}{2}}^{\frac{\pi}{2}} d\tau\int_1^\infty d\rho\,\left[\frac{(aR\tilde\kappa^2\cosh u\,-\,(\tilde\kappa-\rho\sinh u)^2\sinh u )}{2aR\tilde\kappa }\left((\partial_\rho u)^2\,-\,\frac{(\partial_\tau u)^2}{\rho^2-1}\right)\right.\\\nonumber\\\nonumber
&&\left.\,+\,2\gamma\sinh^2 u\,\partial_\rho u\right]\,+\,\frac{2N}{3}\left(\frac{aR\tilde\kappa(\tilde\kappa^2\,+\,\sinh^4 u_+)}{aR\tilde\kappa^2 \cosh u_+\,-\,\sinh u_+(\tilde\kappa-\sinh u_+)^2}\right.\\\nonumber
&&\left.\hspace{4in}+\,u_+\,-\,\sinh u_+\cosh u_+\right)\,.
\eea
The gauge parameter $\gamma$ for the four-form potential must be set to unity, in order to recover the expected result of \cite{Lewkowycz:2013laa} for the symmetric representation in the UV. The UV counterterms in the action ${\rm I}_{\rm D3}$ 
are independent of $\beta$ and do not contribute to entanglement entropy. The terms outside the integral are boundary contributions that arise from evaluating the integrand at the horizon. The $\gamma$-dependent shift also reduces to a horizon contribution:
\be
2\sinh^2u\,\partial_\rho u\,=\,-\partial_\rho(u\,-\sinh u\cosh u)\,.
\ee
Using the expressions in appendix \ref{appa}, this can be reduced to an integral over the domain ${\cal D}$ in Poincar\'e patch, where the integration over time ($x_0$) can be performed analytically and the final integration over the radial coordinate numerically. 
\begin{figure}[h]
\begin{center}
\includegraphics[width=2.7in]{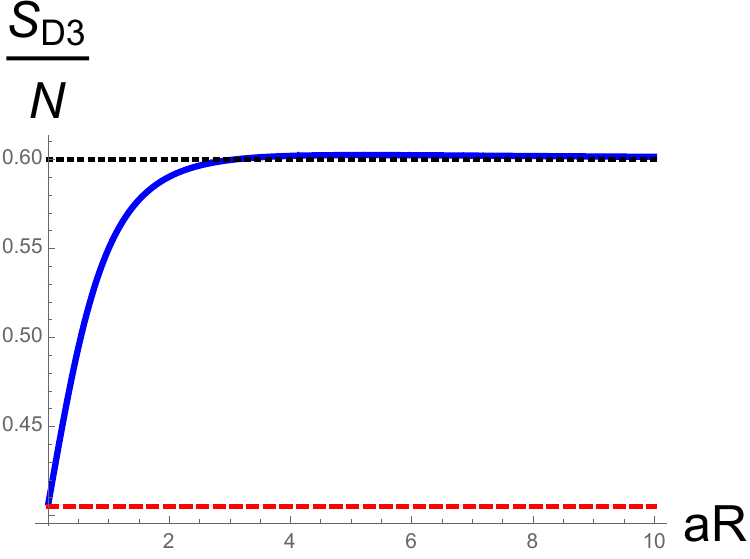}\hspace{0.2in}
\includegraphics[width=2.7in]{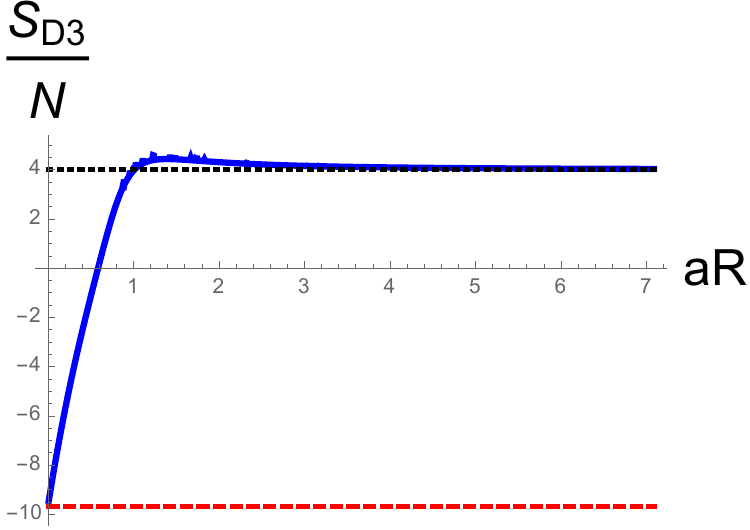}
\end{center}
\caption{\small{{\bf Left:}The D3-brane EE with $a>0$ for $\tilde\kappa=0.9$ (left) increases with increasing $a R$. It appears to overshoot slightly the IR value of $kS_\Box\,=\,\frac{2}{3}N\tilde\kappa$  indicated by the dashed black line. {\bf Right:} The overshoot is clearer at larger $\tilde \kappa$ as shown on right for $\tilde\kappa=6.0$ before it settles to the value for $k$ fundamental quarks.  }}
\label{d3flowaplus}
\end{figure}

\paragraph{{\bf \underline{Positive $a$}}:}The entanglement entropy for $a>0$ solutions exhibits features (figure \ref{d3flowaplus}) that appear counterintuitive at first sight. The EE contribution increases from the UV towards the IR $(aR\gg 1)$. Given the interpretation  of the D3-brane embedding with $a>0$ \cite{Kumar:2016jxy}, and the results of \cite{Lewkowycz:2013laa} this is not really a surprise.  The physical significance of this remains to be understood. We further note that the entanglement entropy rises linearly at $a=0$, whilst the free energy (figures \ref{d3flowaction} and \ref{anegative}) rises quadratically for small $a$. Finally, figure \ref{d3flowaplus} exhibits an overshoot 
before settling down to the large $aR$ value for $k$ fundamental quarks.

\begin{figure}[h]
\begin{center}
\includegraphics[width=2.5in]{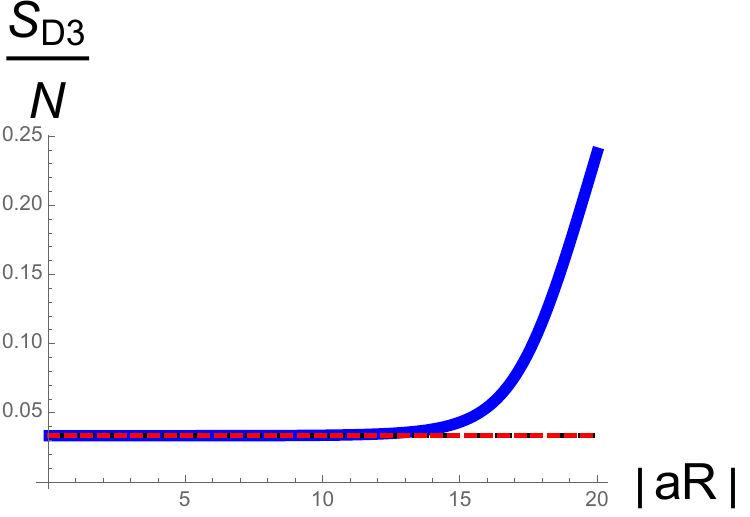}
\hspace{0.3in}
\includegraphics[width=2.5in]{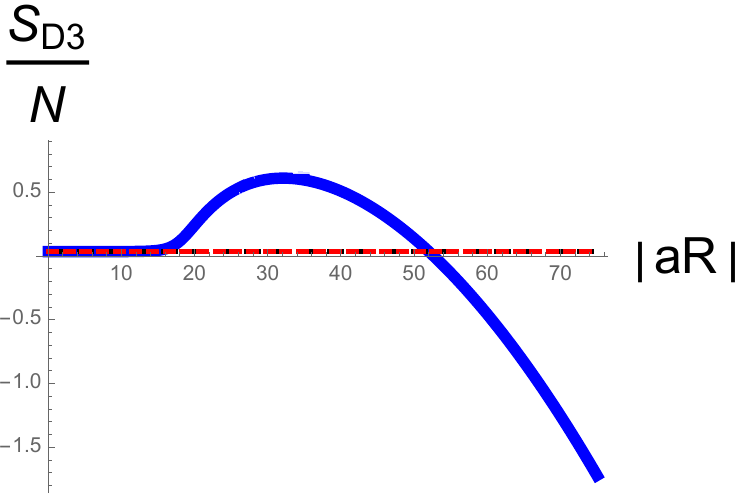}
\end{center}
\caption{\small{The entanglement entropy for $a<0$ solutions with $\tilde\kappa=0.05$. At low values of $|a|R$ it tracks the result for $k$ quarks, departing from it near $|a|R\approx 1/\tilde\kappa$, and finally scaling as $\sim -N a^2$ for large enough $a$. }}
\label{d3eeminus1}
\end{figure}

\paragraph{\underline{\bf Negative $a$:}} Let us now turn to the D3-brane embedding with $a<0$ which represents a heavy quark source in the symmetric representation with $k\sim{\cal O}(N)$ on the Coulomb branch of the ${\cal N}=4$ theory with $SU(N)$ broken to $U(1)\times SU(N-1)$. For small enough $\tilde \kappa$, we can interpret it as a smooth configuration arising from  a collection of $k$ strings ending on a single D3-brane placed at $z\,=\, (\tilde\kappa|a|)^{-1}$. When the size of the entangling sphere on the boundary is small, i.e. $|a| R \ll 1$, the entanglement entropy should match that of $k$ quarks (bundle of coincident strings). In the AdS dual picture, this is due to the fact that the expanded D3-brane remains hidden behind the hyperbolic black hole horizon and we have  $k$ F-strings stretching from the boundary to the hyperbolic horizon. In Poincar\'e patch, this is simply the geometrical statement (depicted in figure \ref{d3intersection}) that the D3-brane embedding intersects the  surface $ z^2 \,+\, r^2 \,=\, R^2$ at $z\approx R$ when $\tilde\kappa \ll 1$ and $|aR| \ll 1/\tilde\kappa$. 
As the size of the region is increased smoothly and $|a R| \approx 1/\tilde\kappa$, the expanded portion of the D3-brane at $z\,=\,1/(\tilde \kappa |a|)$, enters the domain $z^2 + r^2 < R^2$.  At this point the entanglement entropy should exhibit a (smooth) crossover to qualitatively different behaviour which is eventually completely determined by the Coulomb branch VEV. This is precisely what we see in figure \ref{d3eeminus1}.
For large enough $\tilde\kappa$ (greater than a critical value $\tilde\kappa_c\approx 0.6$), the non-monotonic feature  disappears and the EE decreases monotonically.

As in the case of the free energy for $a<0$ solutions, we analyse the integrand in eq.\eqref{d3eeintegral}  and find that the entanglement entropy exhibits the large $|a|$ asymptotic behaviour:
\be
S_{\rm D3}\big|_{aR\gg 1}\,=\,N\left[-\frac{1}{3}(|a| R\tilde\kappa)^2\,-\,c(\tilde\kappa)|a|R\,+\,\frac{2}{3}\ln\left(2 |a| R\tilde\kappa\right)\right]\,.
\ee
We now have both quadratic and logarithmic contributions, and surprisingly, a term linear in $|a|$. The interpretation of the linear term is unclear as we have only determined its coefficient numerically for different values of $\tilde\kappa$. One observation we can make is that the linear term is not simply a Coulomb branch effect as it is not a function of the VEV $\tilde\kappa |a|$ (the separation of the D3-brane from the stack), unlike the other two terms; it also depends nontrivially on $\tilde\kappa$.
\subsection{Comparison with $\langle{\cal O}_{F^2}\rangle$}
\begin{figure}[h]
\begin{center}
\includegraphics[width=2.5in]{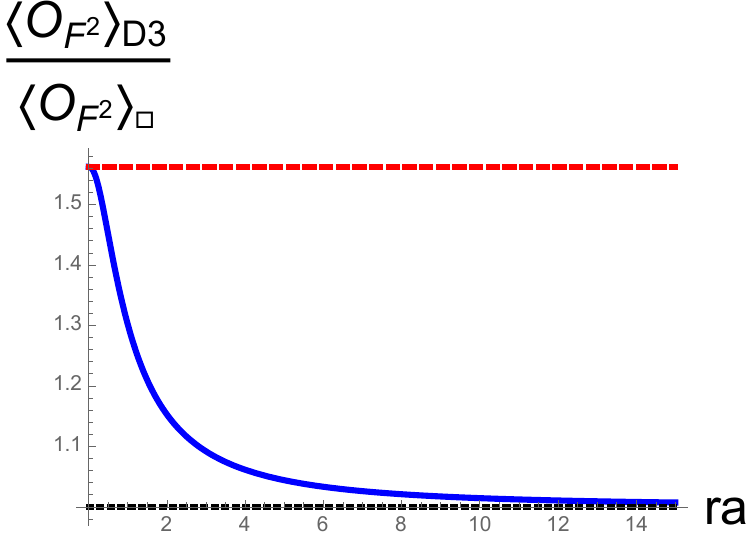}
\hspace{0.3in}
\includegraphics[width=2.5in]{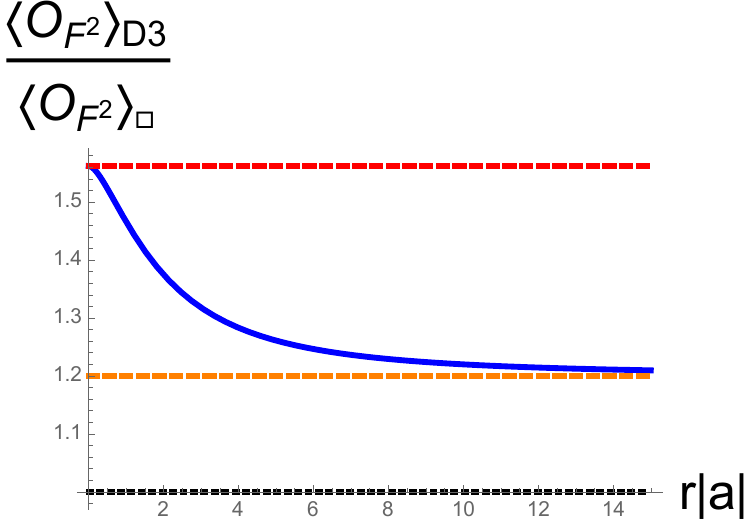}
\end{center}
\caption{\small{ The strength of $\langle{\cal O}_{F^2}\rangle$ sourced by the D3-brane impurity for $a>0$ (left) and $a<0$ (right) with $\tilde\kappa=1.2$. For $a>0$ it interpolates between the value for the symmetric representation ${\cal S}_k$ (dashed red) and $k$ fundamental quarks (dashed black). When $a<0$, the large $r$ value of the ratio $\langle{\cal O}_{F^2}\rangle_{\rm D3}/\langle{\cal O}_{F^2}\rangle_{\rm \Box}\,=\,\tilde\kappa$ (dashed orange), which can be bigger or smaller than unity.}}
\label{monotonic}
\end{figure}

The expectation value of the marginal field theory operator ${\cal O}_{F^2}$ dual to the dilaton reflects the strength of the source.  This expectation value was computed for the D3-brane embedding with $a>0$  in \cite{Kumar:2016jxy} and the strength found to decrease monotonically with distance from the source\footnote{The result for the conformal $(a=0)$ D3-brane embedding was first obtained in \cite{Fiol:2012sg}.}:
\bea
\langle{\cal O}_{F^2}\rangle(r)\,\to&& \frac{\sqrt 2}{4\pi}\,\frac{\tilde\kappa\sqrt{1+\tilde\kappa^2}}{r^4}\,,\qquad 0<ar\,\ll 1\,,\\\nonumber\\\nonumber
\to && \frac{\sqrt 2}{4\pi}\,\frac{\tilde\kappa}{r^4}\,,\qquad ar \gg 1\,.
\eea
Repeating the exercise for the embeddings with $a<0$, we obtain,
\bea
\langle{\cal O}_{F^2}\rangle(r)\,=\,\frac{3\sqrt 2}{16\pi\,r^4}&&\int_0^{\frac{1}{|a|r\,\tilde\kappa }} dy\, (1\,-\, |a|r\,\tilde\kappa y )\, y\,\times\label{trfsquared}\\\nonumber\\\nonumber
&&\left[\left(y^2+\left(1\,-\,\tfrac{\tilde\kappa y }{1\,-\,|a|r\,\tilde\kappa y }\right)^2\right)^{-\frac{5}{2}}\,-\,
\left(y^2+\left(1+\tfrac{\tilde\kappa y }{1-|a|r\tilde\kappa y }\right)^2\right)^{-\frac{5}{2}}\right]\,.
\eea
This expression makes clear that $r^4 \langle{\cal O}_{F^2}\rangle $ is a function of the dimensionless combination $|a| r$. It was obtained by rescaling the integration variable $z$ in eq.\eqref{dild3}, the  radial AdS coordinate,  and defining $y = z/r$.
Using this we find that the asymptotic values for the VEV of the operator for $a<0$ are,
\bea
\langle{\cal O}_{F^2}\rangle(r)\,\to&& \frac{\sqrt 2}{4\pi}\,\frac{\tilde\kappa\sqrt{1+\tilde\kappa^2}}{r^4}\,,\qquad |a|r\,\ll 1\,,\\\nonumber\\\nonumber
\to && \frac{\sqrt 2}{4\pi}\,\frac{\tilde\kappa^2}{r^4}\,,\qquad |a|r \gg 1\,.
\eea
The large $r$ asymptotics is completely controlled by the location of the displaced Coulomb branch D3-brane {\rm with} the flux on it due to the Wilson line or heavy quark probe. In the limit, $|a|r \gg 1$, the  integrand in \eqref{trfsquared}  behaves like a Dirac $\delta$-function and receives all its contributions from a region very close to the location of the Coulomb branch brane (see appendix \ref{integral} for details).
Once again, the strength of the source decreases with increasing distance and as shown in figure \ref{monotonic} it is a monotonic function of $|a|r$.

\section{Discussion and Summary}
The calculations presented in this paper, coupled with the observations in earlier work \cite{Kumar:2016jxy}, raise several questions. We discuss these below.

The worldline theory for the degrees of freedom on a heavy quark impurity in the representation ${\cal A}_k$ (or ${\cal S}_k$) is given by a fermionic (or bosonic) quantum mechanics interacting with  the ${\cal N}=4$ degrees of freedom \cite{passerini}:
\be
{I}_{\rm imp}\,=\,{ I}_{{\cal N}=4}\,+\,\int dt
\left[i\chi^\dagger_m\partial_t\chi^m\,+\, \chi^\dagger_m\left(A_0\,+\,
\hat n^{J}\phi^J\right)^m_n\chi^n\,+\,\mu(\chi^\dagger_m\chi^m-k)\right]\,,\label{qm}
\ee
where $\mu$ is a Lagrange multiplier which enforces the constraint that the fermion or boson number equals $k$. The 
$\{\chi^m\}$ are $N$ flavours (boson or fermion) with $m=1,2,\ldots N$, transforming in the fundamental representation of the $SU(N)$ gauge group. $A_0$ and $\{\phi^I\}$ are, respectively, the temporal component of the gauge field and the six adjoint scalars of the ${\cal N}=4$ theory.  For the superconformal circular (Euclidean)  Wilson loop, the combination $\left(A_0\,+\,\hat n^{J}\phi^J\right)$ can be integrated out \cite{muck, Sachdev:2010uj}  to obtain a bilocal  quartic fermion or scalar interaction:
\be
{ I}_{\rm imp}\,=\,\int_0^\beta d\tau\left(\chi^\dagger_m\partial_\tau\chi^m\,+\frac{\lambda}{8 N\pi^2\beta^2}\int_0^\beta d\tau^\prime\chi^\dagger_m(\tau)\chi^n(\tau)\bar \chi_n(\tau^\prime)\chi^m(\tau^\prime)\right)\,.
\ee
The deformations we have considered in this paper should, in principle, be viewed as deformations of the quantum mechanics \eqref{qm}. Specifically, the D3-brane solution, whose UV description is the Wilson loop in representation ${\cal S}_k$ associated to the bosonic version of the quantum mechanics,  the deformation in question can be naively interpreted as being due to the VEV  of a dimension one operator $\sim \chi^\dagger(\hat n^J\phi^J)\chi$ \cite{Kumar:2016jxy}. This is a singlet under  spatial $SO(3)$ rotations and the $SO(5)_R$ subgroup of the $R$-symmetry left unbroken by internal orientation  (choice of $\hat n$) of the BPS Wilson line. 
The fluctuation analysis in \cite{Kumar:2016jxy} and \cite{Faraggi:2011bb} confirms the existence of such a dimension one operator in the BPS spectrum of the conformal D3-brane embedding dual to the symmetric Wilson loop.
In the fermionic case which corresponds to the  IR description of the D5-brane  impurity, the deformation is by an irrelevant operator of dimension four $\sim \chi^\dagger \left(\hat n^JD_\alpha \phi^J\right)^2 \chi$ \cite{Harrison:2011fs}. 

It would be extremely interesting to understand how the flows indicated by the brane embeddings  emerge from deformations of the impurity quantum mechanics discussed above. The naive interpretation of the D3-brane deformation as a VEV of a dimension one operator \cite{Kumar:2016jxy} in the impurity theory is particularly puzzling, as it would imply spontaneous breaking of conformal invariance in the associated quantum mechanics (which should not be possible). This is plausible for the case $a<0$ which corresponds to a Coulomb branch configuration. In this situation the deformation gets related to a scalar VEV in the ambient four dimensional theory, which feeds into the impurity quantum mechanics. However, this does not explain what happens  for $a>0$ deformations.  It is worth noting that the D5-brane solution  also appears to be triggered in the UV by the VEV of a dimension one operator \cite{Kumar:2016jxy}.

By explicit evaluation we have seen that the free energy $-\ln Z_{\cal H}$ on ${\cal H}\simeq S^1\times H^3$ for each of the nonconformal impurities (both D5- and D3-brane) is a monotonically increasing function of the size of the deformation parameter. This is the behaviour expected for the relative entropy of excited states, relative to the thermal one. On the other hand, the change in the entanglement entropy of a region of radius $R$ (in flat space) due to the D5-brane impurity and the $a<0$  D3-brane defect, is non-monotonic for small $R$, eventually becoming a decreasing function of $R$ for large $R$. This statement, however, does not apply to the $a>0$ D3-brane solution for which the jump in EE, while always bounded,  {\em increases} non-monotonically  and saturates at a finite value in the IR. 

The fact that the jump in EE due to the (pointlike) defects is non-decreasing or non-monotonic, while puzzling, is not immediately in conflict with the $g$-theorem  \cite{Affleck:1991tk, Friedan:2003yc} and its holographic version \cite{Fujita:2011fp, Casini:2016fgb}. The latter apply to CFTs in 1+1 dimensions with a boundary impurity or to CFTs in $d$ dimensions with a $d-1$ dimensional boundary. This includes the Kondo model where an effective 1+1 dimensional description is obtained after reducing to the $s$-wave modes \cite{Affleck:1995ge}. Evidently, this is not the case for our problem where the pointlike impurity is placed in an ambient $3+1$ dimensional CFT at large-$N$, with an AdS$_5$ gravity dual (without degrees of freedom confined to an AdS$_3$ subspace as in \cite{Erdmenger:2013dpa}).

Overall, it would clearly be very interesting to understand the physical reason behind the very different behaviours of entanglement entropy for the different types of sources and how these qualitative features of entanglement relate to the strength of the sources as indicated by the long range falloff of the fields coupled to them.

Technically, the calculation of the gravitational entropy contribution from the D3-brane probes involved a new aspect not encountered previously, namely, the role of RR potentials and their inherent gauge ambiguity. The choice of gauge was fixed by matching to the result of \cite{Lewkowycz:2013nqa} for entanglement entropy of the symmetric representation source in the absence of any deformation. Given that this is  crucial for obtaining the correct result via the replica method, a first principles understanding of the gauge choice would be desirable.

\acknowledgments
We have benefitted from many enjoyable discussions with Costas Bachas, Justin David, Andy O'Bannon, Carlos N\'u\~nez, David Tong and Dan Thompson. SPK would also like to thank the Galileo Galilei Institute for Theoretical Physics (GGI) for hospitality and providing a stimulating atmosphere, and INFN for partial support, while this work was being completed during the program ``New Developments in AdS$_3$/CFT$_2$ Holography".
This work was supported by STFC grants ST/L000369/1, ST/P00055X/1 and ST/K5023761/1.
%\startappendix
\appendix
\section{Transformations for D3-brane embedding}
\label{appa}
For the D3-brane non-conformal embedding, we need to evaluate the action and its derivative with respect to $\beta$ in the hyperbolic AdS  geometry.  The embedding \eqref{hypadsbps} implicitly specifies  $u = u(\tau,\rho)$. To calculate the action and the entanglement entropy, we require the derivatives $\partial_\tau u$ and $\partial_\rho u$ as  functions of the hyperbolic AdS coordinates $(\tau, \rho, u)$ and  Poincar\'e patch coordinates $(x_0, z, r )$. We first use the transformations \eqref{poinctohyp} (after Wick rotation to Euclidean signature) and the BPS solution \eqref{d3bps} to evaluate the derivatives in the D3-brane action:
\bea
\frac{\partial u}{\partial \rho}\,=\,\frac{-\frac{\rho\cos\tau}{\sqrt{\rho^2-1}}-\cosh u\,+\frac{aR\,\tilde\kappa^2\sinh u}{(\tilde\kappa-\rho\sinh u)^2}}{\rho\left(\sinh u-\frac{aR\,\tilde\kappa^2\cosh u}{(\tilde\kappa-\rho\sinh u)^2}\right)}\,,
\quad
\frac{\partial u}{\partial \tau}\,=\,\frac{\sqrt{\rho^2-1}\sin\tau}{\rho\left(\sinh u-\frac{aR\,\tilde\kappa^2\cosh u}{(\tilde\kappa-\rho\sinh u)^2}\right)}\,.
\eea
Next, we note that the combination of derivatives of $u$ that appears in the DBI portion of the D3-brane action simplifies considerably:
\bea
1\,+\,\rho^2(\rho^2-1)(\partial_\rho u)^2\,+\,\frac{\rho^2(\partial_\tau u)^2}{\rho^2-1}\,=\,\frac{(aR)^2\tilde\kappa^2\left(\tilde\kappa^2\,+\,\rho^4\sinh^4u\right)}{\left[aR\tilde\kappa^2\cosh u\,-\,(\tilde\kappa-\rho\sinh u)^2\sinh u\right]^2}
\eea
Then the on-shell action in hyerbolic AdS space with $\beta=2\pi$ is 
\bea
&&{\rm I}_{\rm D3}(2\pi)\,=\,\frac{2N}{\pi}\int_{-\frac{\pi}{2}}^{\frac{\pi}{2}}d\tau\int_1^\infty d\rho \left[\frac{aR\tilde\kappa\left(\tilde\kappa^2\,+\,\rho^4\sinh^4u\right)}{aR\tilde\kappa^2\cosh u\,-\,(\tilde\kappa-\rho\sinh u)^2\sinh u}\,+\right.\label{d3actioneval}\\\nonumber\\\nonumber
&&\left. + \rho^2(\rho^2-1)\sinh^2u\,\partial_\rho u\,+\,\rho(u-\sinh u\cosh u)\right]\,+\,
{\rm I}_{\rm UV}\,.
\eea
Evaluating the integral in hyperbolic space coordinates is unwieldy since $u(\rho,\tau)$ is a complicated function. Instead, we translate back to Poincar\'e patch coordinates. The Jacobian for this transformation, (evaluated on the D3-brane worldvolume) is:
\be
\left|\frac{\partial\rho}{\partial z}\frac{\partial \tau}{\partial x_0}\,-\,\frac{\partial\rho}{\partial x_0}\frac{\partial \tau}{\partial z}\right|\,=\,\frac{1}{z^2}\,\frac{1}{(aR)\,\tilde\kappa^2}\left((aR)\,\tilde\kappa^2\cosh u\,-\,\sinh u(\tilde\kappa -\rho\sinh u)^2\right)
\ee
The hyperbolic AdS coordinates can be written in terms of Poincar\'e patch variables:
\bea
&&\rho\,=\,\frac{\sqrt{r(z)^4\,+\,2\,r(z)^2(z^2+x_0^2-R^2)\,+\,(R^2+x_0^2+z^2)^2}}{2\,z \,R}\,,\\\nonumber\\\nonumber
&&\sin \tau\,=\,\frac{2\,x_0\,R}{\sqrt{r(z)^4\,+\,2\,r(z)^2(z^2+x_0^2-R^2)\,+\,(R^2+x_0^2+z^2)^2\,-\,4\,z^2\,R^2}}\,,\\\nonumber\\\nonumber
&&\sinh u\,=\,\frac{2\,r(z) \,R}{\sqrt{r(z)^4\,+\,2\,r(z)^2(z^2+x_0^2-R^2)\,+\,(R^2+x_0^2+z^2)^2}}\,,
\eea
where $r(z)\,=\,\tilde\kappa z/(1+a\tilde\kappa z)$.  We may now simplify the individual terms in the integrand in \eqref{d3actioneval} quite substantially. The Jacobian and the DBI piece combine to yield:
\be
{\cal I}_1\,=\,d\tau\,d\rho\,\frac{aR\tilde\kappa\left(\tilde\kappa^2\,+\,\rho^4\sinh^4u\right)}{aR\tilde\kappa^2\cosh u\,-\,(\tilde\kappa-\rho\sinh u)^2\sinh u}\,=\, dx_0\,dz\,\left(\frac{\tilde\kappa}{z^2}\,+\,\frac{r(z)^4}{z^6\,\tilde\kappa}\right)\,.
\ee
The Wess-Zumino terms can be combined to yield a piece which is a total derivative:
\bea
&&\rho^2(\rho^2-1)\sinh^2 u\,\partial_\rho u\,+\, \rho(u\,-\,\sinh u\cosh u)\\\nonumber
&&\hspace{2in}=\,
\rho^4\sinh^2 u\,\partial_\rho u\,+\,\frac{1}{2}\,\frac{\partial}{\partial \rho}\left[\rho^2\left(u\,-\,\sinh u\cosh u\right)\right]\,.
\eea
Transforming the first of these two to Poincar\'e patch variables we find:
\bea
&&{\cal I}_2\,=\,d\tau\,d\rho\,\rho^4\sinh^2 u\,\partial_\rho u\,\,=\,
dx_0\,dz\left[-\frac{r(z)^3}{z^5}\,+\,\frac{r(z)^2}{z^5 a\kappa^2 }\left(\tilde\kappa-\frac{r(z)}{z}\right)^2\times\right.\\\nonumber
&&\left.\left(\frac{\left\{(x_0^2+R^2)^2-z^4+2 r(z)^2(x_0^2-R^2)+r(z)^4\right\}}{r^4+2r(z)^2(x_0^2+z^2-R^2)+(x_0^2+z^2)^2+2R^2(x_0^2-z^2)+R^4}\right)\right]
\eea
Finally the total derivative contribution evaluates to
\be
{\cal I}_3\,=\,-N \left(u_+\,-\,\sinh u_+\,\cosh u_+\right)\,,\qquad \sinh u_+\,=\,\frac{\tilde\kappa \,\cosh u_+}{\tilde\kappa\, aR\,+\, \cosh u_+}\,,\label{uplus}
\ee
where $u_+$ is the value at the hyperbolic horizon $\rho=1$, and the contribution from the boundary at $\rho\to\infty$ is vanishingly small. The complete action for the solution can now be written as the sum of these different terms:
\be
{\rm I}_{\rm D3}(2\pi)\,=\,\frac{2N}{\pi} \int\int_{\cal D} \left({\cal I}_1\,+\,{\cal I}_2\right)\,-\,N\left(u_+\,-\,\sinh u_+\,\cosh u_+\right)\,-\,4N\tilde\kappa\frac{R}{\pi\epsilon}\,. 
\ee
where, the last term is the UV counterterm, and,
\bea
&&\int\int_{\cal D} \left({\cal I}_1\,+\,{\cal I}_2\right)\,=\,\int_{\epsilon}^{z_{\rm max}} dz
\int_{-\sqrt{R^2-r^2-z^2}}^{\sqrt{R^2-r^2-z^2}}dx_0\,\left[\frac{\tilde \kappa}{z^2}\,-\,\frac{a \tilde \kappa^4}{z(1+az\tilde\kappa)^4}\right.\\\nonumber\\\nonumber
&&\left. +\,\frac{a\tilde\kappa^4}{z(1+az\tilde\kappa)^4}
\left\{\frac{\left((x_0^2+R^2)^2-z^4+2 r(z)^2(x_0^2-R^2)+r(z)^4\right)}{r(z)^4+2r(z)^2(x_0^2+z^2-R^2)+(x_0^2+z^2)^2+2R^2(x_0^2-z^2)+R^4}\right\}\right]\,.
\eea
 \section{Evaluation of $\langle{\cal O}_F^2\rangle_{\rm D3}$}
 \label{integral}
Following the analysis presented in \cite{Kumar:2016jxy} and setting $a =- |a|$, we have
\bea
\langle{\cal O}_F^2\rangle_{\rm D3}\,=\,\frac{3\sqrt 2}{16\pi\,r}&&\int_0^{|a \tilde\kappa|^{-1} }dz\,(1-|a|\tilde\kappa z)\,z\,\times\label{dild3}\\\nonumber\\\nonumber
&&\left[\left(z^2+\left(r\,-\,\tfrac{\tilde\kappa z}{1-|a|\tilde\kappa z}\right)^2\right)^{-\frac{5}{2}}\,-\,\left(z^2+\left(r\,+\,\tfrac{\tilde\kappa z}{1-|a|\tilde\kappa z}\right)^2\right)^{-\frac{5}{2}}\right]\,.
\eea
The $z$-integration is cut off at $z=1/|a|\tilde\kappa$ where the embedding terminates in the blown up Coulomb branch D3-brane. We can then define the rescaled variable $x\,=\,|a|\tilde\kappa z$ which yields:
\bea
\langle{\cal O}_F^2\rangle_{\rm D3}\,=\,&&\frac{3\sqrt 2}{16\pi r^6\tilde\kappa^2 a^2}\int_0^1\,dx\,(1-x)\,x\,
\times\\\nonumber\\\nonumber
&&\left[\left(\frac{x^2}{|a\tilde\kappa r|^2}\,+\,\left(1\,-\,\tfrac{x/|a r|}{1-x}\right)^2\right)^{-\frac{5}{2}}\,-\,\left(\frac{x^2}{|a\tilde\kappa r|^2}\,+\,\left(1\,+\,\tfrac{x/|a r|}{1-x}\right)^2\right)^{-\frac{5}{2}}\right]\,.
\eea 
Since $x$ is bounded between $0$ and $1$, in the large $r$ limit we need to examine the function 
\be
\lim_{\epsilon\to 0}\frac{1}{(y^2\,+\,\epsilon^2)^{n } }\,.
\ee
This is sharply peaked at $x=0$. Integrating across any finite interval containing $y=0$, we obtain
\be
\lim_{\epsilon\to 0}\int_{-b}^bdy\,\frac{1}{(y^2\,+\,\epsilon^2)^{n } }\,=\,\frac{\sqrt\pi \Gamma\left(n-\tfrac{1}{2}\right)}{\epsilon^{2n-1}\,\Gamma(n)}\,.
\ee
Therefore we make the replacement,
\be
\frac{1}{(y^2\,+\,\epsilon^2)^{n } }\,\to\,\frac{\sqrt\pi \Gamma\left(n-\tfrac{1}{2}\right)}{\epsilon^{2n-1}\,\Gamma(n)}\,\delta(y)\,.
\ee
Only the first of  the two terms in the expression for $\langle{\cal O}_F^2\rangle_{\rm D3}$ yields a $\delta-$function with support in the interval $0\leq x\leq 1$. Thus, assuming $|a|r\gg 1$,
\be
\langle{\cal O}_F^2\rangle_{\rm D3}\,=\,\frac{\sqrt 2\,\tilde\kappa^2 a^2}{4\pi r^2} \int_0^1dx (1-x) x^{-4} \delta\left(\frac{1}{x}-\frac{1}{a|r|(1-x)}\right)\,=\,
\frac{\sqrt 2\,\tilde\kappa^2 }{4\pi\, r^4}\,,
\ee
where the $\delta$-function has support at $x= |a|r/(1+|a|r)$.

\end{document}